\documentclass[twocolumn,
aps,
prb,
amsmath,
amssymb,
floatfix,
noeprint,
reprint]{revtex4-2}

\usepackage{graphicx,xcolor,mathtools,bm,MnSymbol}
\usepackage[normalem]{ulem}
\usepackage{soul}    
\usepackage{physics}
\usepackage{verbatim}
\usepackage{verbatim}
\usepackage{esint}

\def \beq {\begin{eqnarray}}
\def \eeq {\end{eqnarray}}

\definecolor{DarkRed}{RGB}{100,0,0}

\definecolor{DarkGreen}{RGB}{0,100,0}

\usepackage[bookmarks=false,linkcolor=blue,urlcolor=blue,colorlinks,citecolor=blue]{hyperref}
\usepackage[all]{hypcap}
\newcommand{\sign}{\text{sgn}}
\usepackage[english]{babel}
\usepackage[babel,kerning=true,spacing=true]{microtype}

\begin{document}

\title{Weakly damped bosons and precursor gap in the vicinity of an antiferromagnetic metallic transition}

\author{Ori Grossman}
\email{ori.grossman@weizmann.ac.il}
\author{Erez Berg}
\email{erez.berg@weizmann.ac.il}
\affiliation{Department of Condensed Matter Physics, Weizmann Institute of Science, Rehovot, 76100, Israel.}

\begin{abstract}
 We study the electronic spectral function of a metal in the vicinity of an antiferromagnetic (AFM) quantum critical point, focusing on a situation where the bare bandwidth of the spin fluctuations is significantly smaller than the Fermi energy. 
 In this limit, we identify a range of energies where the fermionic quasiparticles near the ``hot spots'' on the Fermi surface are strongly scattered by the quantum critical fluctuations, whereas the damping of the AFM fluctuations by the electrons is negligible. 
Within a one-loop approximation, there is a parameter range where the $T=0$ spectral function at the hot spots has a ``precursor gap'' feature, with a local maximum at a finite frequency. 
However, the ratio of the bare spin wave velocity to the Fermi velocity required to obtain a precursor gap is probably too small to explain experiments in the electron-doped cuprate superconductors~\cite{He2019}. 
At lower frequencies, the Landau damping of the AFM fluctuations becomes important, and the electronic spectral function has the familiar ${\omega}^{-1/2}$ singularity. 
Our one-loop perturbative results are supported by a numerical Monte Carlo simulation of electrons coupled to an undamped, nearly-critical AFM mode. 
\end{abstract}

\maketitle

\section{Introduction} 

Continuous antiferromagnetic (AFM) transitions are common among many strongly correlated metals, including heavy fermions, cuprates, and iron-based superconductors. Despite decades of intense research \cite{Hertz1976,Millis1993,Abanov2000,Abanov2003,Abanov2004,Metlitski2010,Bergeron2012,Efetov2013,Lee2013,Varma2015,Varma2015a,Schlief2017,Lee2018}, the subtle interplay between low-energy magnetic fluctuations and electronic quasi-particles keeps producing surprises. A particularly important question is whether the transition from a metal to a metallic antiferromagnet occurs directly, or are there intermediate phases that separate the simple Fermi liquid (FL) and the AFM metal.

The electron-doped cuprates \cite{Armitage2010} exhibit a broad regime of strong AFM fluctuations, with long AFM correlation lengths \cite{motoyama2007spin} and clear evidence for hot spots in the electronic spectrum \cite{Armitage2002}. A recent angle-resolved photoemission spectroscopy (ARPES) study \cite{He2019} in Nd$_{2-x}$Ce$_x$CuO$_4$ revealed that, surprisingly, a gap-like feature appears in the electronic spectrum at the AFM hot spots already at doping levels where  long-range antiferromagnetic order is absent. The experiment was performed at low temperature, such that thermal AFM fluctuations are not likely to play an important role. These results are particularly intriguing, since a Fermi surface (FS) reconstruction at $T=0$ without long-range order would imply that the ground state violates Luttinger's theorem~\cite{Luttinger1960,Oshikawa2000}, and is hence not a simple FL \cite{Senthil2003,Scheurer2018}.

Conversely, the experimental results raise the question whether such a ``precursor gap'' feature may appear in the non-AFM side of a more conventional, direct quantum phase transition from an AFM to a simple metal~\footnote{In this discussion, we ignore the possibility of superconductivity in the vicinity of the quantum critical point.}. Such a precursor gap in the electronic spectrum is known to arise in the magnetically disordered phase due to \emph{thermal} (static) fluctuations \cite{Schmalian1999,Ye2023}. Whether a similar feature can appear upon approaching the AFM quantum critical critical point (QCP) at zero temperature is unclear. At asymptotically low frequencies, the conventional one-loop treatment~\cite{Sachdev2011} predicts that the fermion spectral function diverges as $|\omega|^{-1/2}$ in the limit of small frequency, with no local maximum at $\omega>0$.

In this work, we examine the electronic spectral function of a nearly-AFM metal, described by the spin-fermion model \cite{Schmalian1999}. We focus on the case where the bare speed characterizing the magnetic fluctuations, $v_s$, is significantly smaller than the Fermi velocity at the hot spots, $v_F$.
In the limit $v_s\ll v_F$, we show that there is a range of energies where the effects of the Landau damping of the spin fluctuations by the electrons are small. To address this regime, we study the spectral properties of electrons coupled to \emph{undamped} AFM fluctuations. The electron spectral function is computed either perturbatively in the coupling between electrons and AFM fluctuations, or numerically, within the ``quenched approximation'' \cite{Hamber1981,Meszena2016}, where the electron self-energy is computed to all orders, but the feedback of the electrons on the dynamics of the spin fluctuations is neglected.
At sufficiently low energies, the feedback effects of the electrons on the AFM fluactuations (Landau damping) become significant even for $v_s\ll v_F$, and the system crosses over to the more conventional regime where the electronic and the AFM degrees of freedom have to be taken into account on equal footing.

In the undamped case, the one-loop $T=0$ electronic spectral function in the disordered side of the QCP can exhibit a ``precursor gap'' feature at the hot spots, with a local maximum at finite frequency. 
However, within the one-loop approximation, obtaining a precursor gap at the hot spots requires quite small values of $v_s/v_F$ -- about an order of magnitude smaller than the ratio in the electron-doped cuprates. Conventional quantum AFM fluctuations are therefore unlikely to explain the low-temperature precursor gap observed in this system~\cite{He2019}.

To go beyond the one loop level, we perform Monte Carlo simulations of the imaginary-time Green's function of fermions coupled to undampled AFM fluctuations. The numerical results are found to agree qualitatively with the one-loop calculations up to $\xi\approx 5a$, where $\xi$ is the AFM correlation length and $a$ is the lattice spacing.

This paper is organized as follows. In Sec. \ref{sec:model} we set up the spin-fermion model for a nearly-antiferromagnetic metal. Sec. \ref{sec:undamped} describes the different energy scales of the model, and identifies the regime where the antiferromagnetic fluctuations can be treated as undamped. In Sec. \ref{sec:oneloop} we describe the electronic spectral function within a one-loop approximation, followed by numerical Monte Carlo results in the quenched approximation, described in Sec. \ref{sec:numerical}. The results are discussed in Sec. \ref{sec:discussion}.

\section{Model}
\label{sec:model}
\begin{figure}[t]
        	\centering\includegraphics[ width=1\columnwidth]{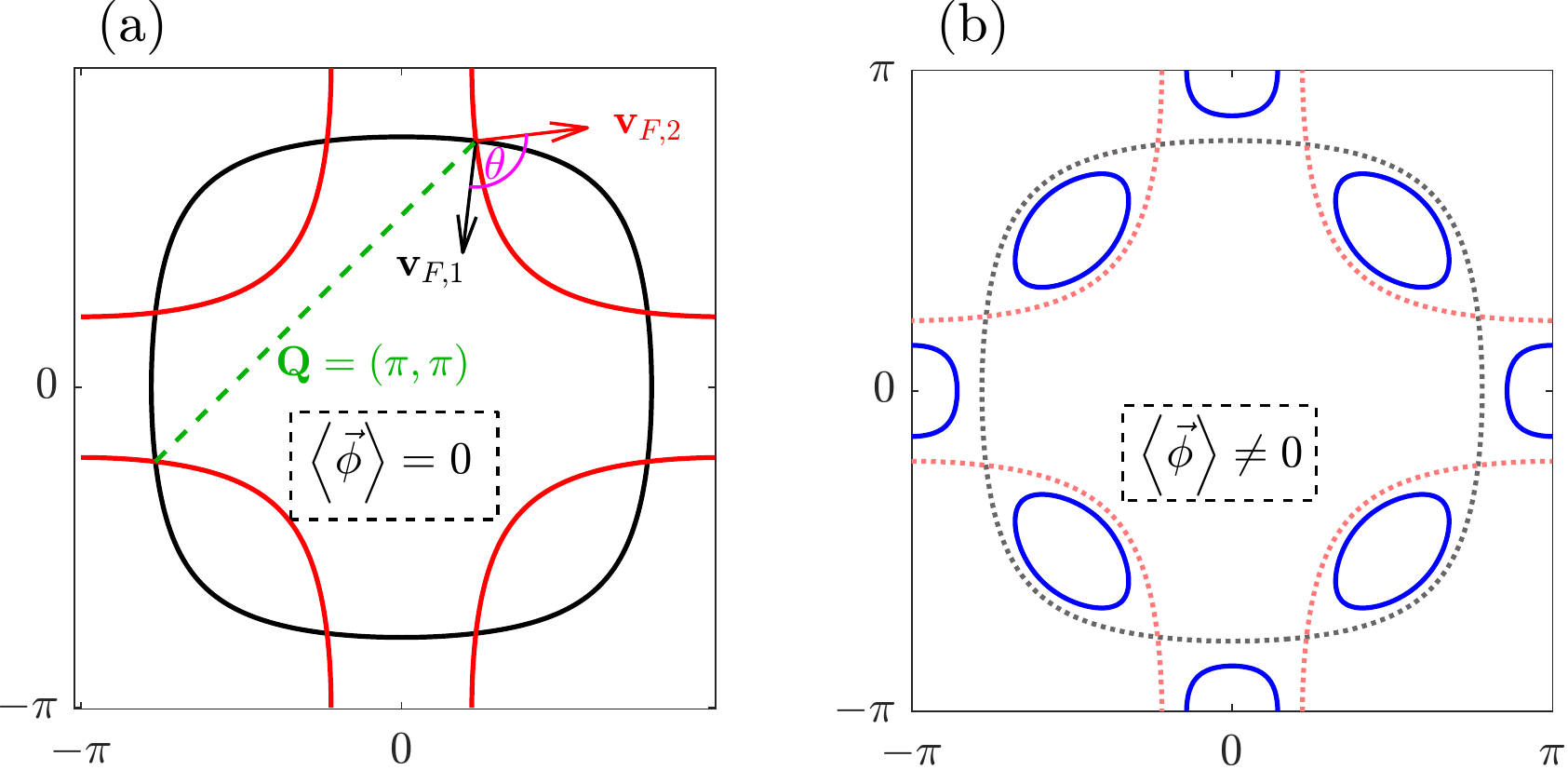}
	\caption{An illustration of the Fermi surface (FS) in both phases. (\textbf{a}) The disordered phase with no expectation value for the bosonic order parameter, where The FS is singly connected.  (\textbf{b}) The ordered phase, where the FS (1, black) and the shifted FS by $\mathbf{Q}=\left(\pi,\pi \right)$ (2, red), are hybridized. The FS is reconstructed  into electron and hole pockets (blue).}
\label{fig:illustration}
\end{figure}

Metallic AFM phase transitions are characterized by hot spots on the FS, at which the quasiparticles can scatter resonantly off the critical spin fluctuations. 
Here, we consider the widely studied spin-fermion model \cite{Schmalian1999}, 
which  captures the key feature 
of the interplay between the fermionic gapless modes and the spin collective modes.
The model is defined on a square lattice, with an action given by $S=S_{\psi}+S_{int}+S_{\phi}$, where
\begin{align}
    S_{\psi}&=\int_{\mathbf{k},\omega} \ \  \sum_{s} \psi^{\dagger}_{\mathbf{k},\omega,s} \left( i\omega -  \epsilon_{\mathbf{k}} \right) \psi_{\mathbf{k},\omega,s} \notag, \\ \quad 
    S_{int}&= g \int_{\omega,\Omega,\mathbf{k},\mathbf{q}} \ \ \sum_{s,s'} \vec{\phi}_{\mathbf{q},\Omega}\cdot \left( \psi^{\dagger}_{\mathbf{k},\omega,s} 
    \vec{\sigma}_{s,s'}
     \psi_{\mathbf{k+q+Q},\omega,s'} +\rm{h.c}  \right), \notag \\ \quad 
    S_{\phi}&=  \int_{\Omega,
    \mathbf{q}} \left[ 
    \left(\frac{\Omega^2}{v_{s}^2} +\mathbf{q}^2 +\xi_{0}^{-2} \right) |{\vec{\phi}_{\mathbf{q},\Omega}}|^2  
    \right]. 
     \label{Eq:model}
\end{align}
    For convenience, we have adopted the short hand notation $ \int_{x_1,x_2... x_n}=\displaystyle  \prod_{i=1,2...n}\int \frac{\textrm{d}x_i}{2\pi}$. 
    Here, $\mathbf{k},\mathbf{q}$ are crystal momenta, $s \in \lbrace\uparrow,\downarrow \rbrace$ denotes spin,
     and $\vec{\sigma}$ is  the vector of Pauli matrices. 
$S_{\psi}$ is the kinetic part of the fermions, whose dispersion is denoted by $\epsilon_{\mathbf{k}}$. 
The FS, determined by the condition $\epsilon_{\mathbf{k}}=0$,  includes pairs of hot spots connected by the ordering vector $\mathbf{Q}=(\pi,\pi)$.
$S_{\phi}$ is the action of the bosonic collective mode $\vec{\phi}$, which is a three-component vector field (we neglect spin-orbit coupling). The bosonic fluctuations are coupled to the spin operator of the fermions by $S_{int}$. 
The bare spin wave velocity is denoted by $v_s$. 
$\xi_0$ is the bare (unrenormalized) correlation length. The actual AFM transition occurs when $\xi_0$ reaches a certain critical value, which we denote by $\xi_c$. 
An illustration of the FS reconstruction in the AFM ordered phase is shown in Fig.~\ref{fig:illustration}.

\section{Undamped boson regime} 
\label{sec:undamped}
We now show that when the bare spin wave velocity $v_s$ is much smaller than the Fermi velocity at the hot spots (which we denote by $v_F$), there is a range of energies where the Landau damping of the spin fluctuations by the electrons is negligible, i.e., the spin fluctuations are underdamped. In the same regime, the scattering of electrons at the hot spots by the AFM fluctuations is significant, as we shall explore in detail next.  

As is well known, the damping of the AFM fluctuations by electrons at the hot spots modifies qualitatively the dynamics of the former, and changes the properties of the quantum critical point. Within a one-loop approximation, the renormalized boson propagator takes the form
\begin{equation}
D\left(\mathbf{q},i\Omega_n\right)= \left( \xi^{-2} +\mathbf{q}^2+\Omega^2/v_{s}^2 +\gamma\left| \Omega \right| \right)^{-1}
\label{Eq:bosonic_propagator}
\end{equation}
where $\xi$ is the renormalized  AFM correlation length such that $\xi^{-2} = \xi_0^{-2}-\xi_c^{-2}$, and the Landau damping coefficient $\gamma$ is given by
\begin{equation}
\gamma=\frac{Ng^2}{\pi v_F^2 \sin \theta}, 
\end{equation}
 where $\theta \in \left[0,\pi \right] $ is the angle between the Fermi velocities at the hot spot (see Fig. \ref{fig:illustration}), and $N$ is the number of hot spot pairs (here $N=4$). At low energies, the Landau damping term dominates the frequency dependence, changing the dynamical critical exponent from $z=1$ to $2$~\footnote{In the ultimate infra-red fixed point, $z$ flows back to 1~\cite{Schlief2017}.}.
 By comparing the two frequency-dependent terms in Eq. \eqref{Eq:bosonic_propagator}, we obtain an energy scale
\begin{equation}
    \Omega_{B}=v_{s}^2\gamma=\frac{Ng^2 v_{s}^2}{\pi v_F^2 \sin \theta}
    \label{Eq:Omega_B_def}
\end{equation}
where the crossover between $z=1$ and $z=2$ occurs. 
For frequencies $\omega \ll \Omega_B$, the Landau-damping term dominates, and the dynamics is overdamped. Interestingly, in the limit where $v_s\ll v_F$, $\Omega_B$ can be made much smaller than the Fermi energy $E_F$. 
In this work, we assume that this is the case, such that there is a frequency window
    $\Omega_B \ll \Omega \ll E_F$ in which the AFM fluctuations are essentially undamped.

Naively, one may expect that at sufficiently high frequency, $\omega\gg \Omega_B$, the fermionic self-energy can be calculated using the undamped form of the bosonic propagator (setting $\gamma$ to zero in Eq.\eqref{Eq:bosonic_propagator}). A careful analysis shows, however, that this approximation is only valid for a more restricted range of frequency, $\Omega_B \frac{v_F}{v_s} \ll \omega \ll E_F$.  
In this regime we find a new energy scale, $\lambda$,  given by
\begin{equation}
   \lambda = \frac{3v_s g^2}{4\pi v_F},
  \label{eq:lambda}
\end{equation}
such that the spectral function at the hot spot is a function of   $\omega$, $\lambda$, and $E_F$. 
Note that $\lambda$ and  $\frac{v_F}{v_s}\Omega_B$ are 
parametrically the same.

\begin{figure}
        	\centering\includegraphics[ width=1\columnwidth]{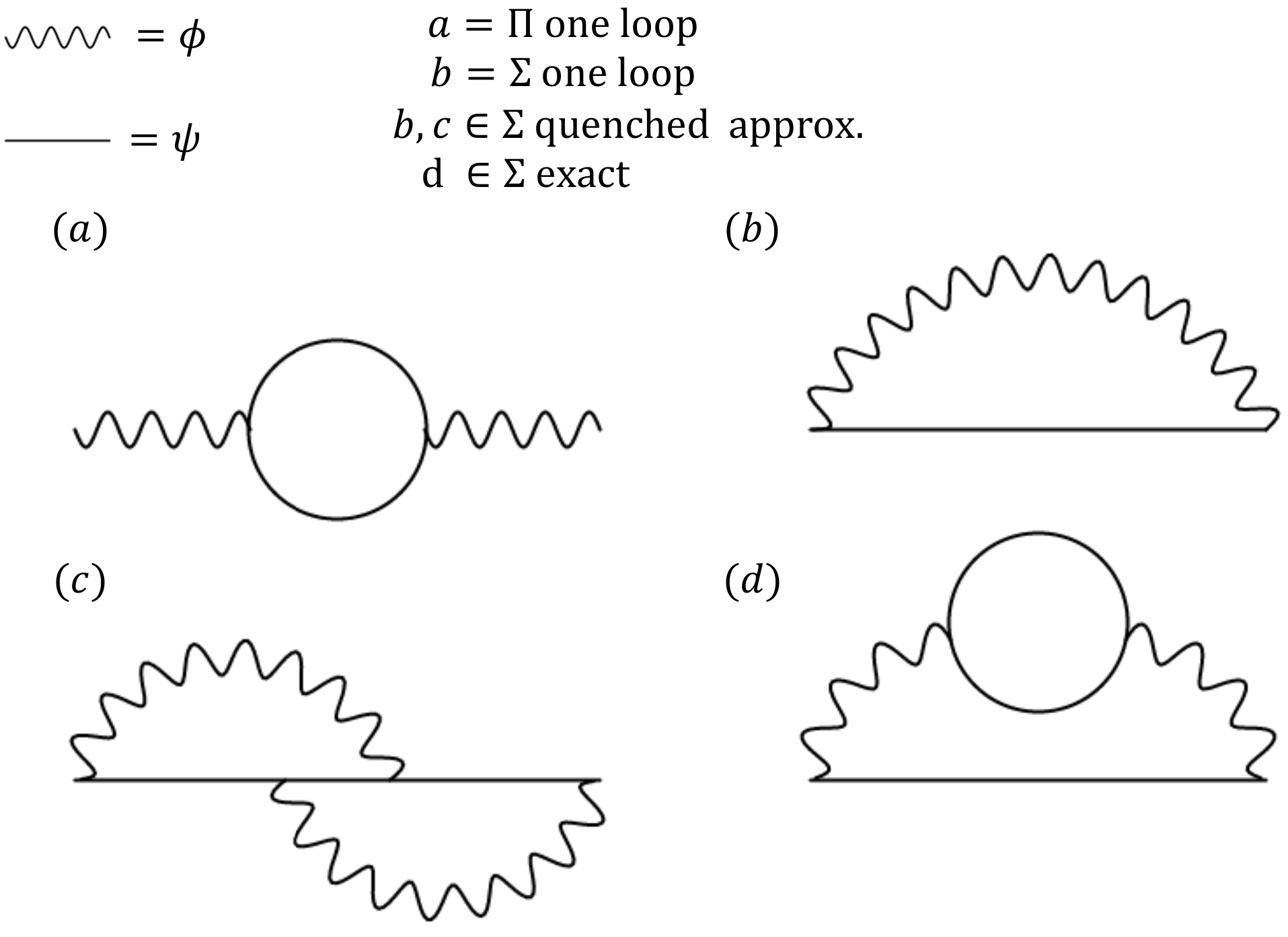} 
	\caption{An illustration of the  diagrams which participate in the quenched approximation, compared to the exact sum.}
\label{fig:diagram_quench}
\end{figure}

\section{One-loop fermion self energy in weakly damped boson regime}
\label{sec:oneloop}
We now examine the fermion self-energy due to scattering off the AFM fluctuations. The leading order self-energy diagram, shown in Fig. \ref{fig:diagram_quench}b, in the vicinity of the hot spots is given by 
\begin{align}
\label{Eq: Sigma_Matsubara}
&\Sigma \left( \delta\mathbf{k+k}_{HS},i\omega \right)=\\ \quad \notag
3&g^2\int_{\mathbf{q},\Omega} G_{0}\left(\mathbf{q+Q+\delta k+ k}_{HS},i\omega+i\Omega\right)D\left(\mathbf{q},i\Omega\right).
\end{align}
where $\mathbf{{k}_{HS}}$ is the hot spot wavevector, and $G_0(\mathbf{k},i\omega) = (i\omega - \epsilon_{\mathbf{k}})^{-1}$ is the bare electron Green's function. The factor 3 of the $SU(2)$ dimension, is due to identical contribution of $\vec{\phi}$ in each direction.

We start by focusing on the $T=0$ spectral function  $A \left( \mathbf{{k}_{HS}},\omega \right)$ within the undamped regime, 
where effectively $\gamma=0$. 
To this end, we consider a linearized model with two bands. The different bands represent the dispersion around  each of the two points which form together a hot spot pair, $\mathbf{{k}_{HS}}$ and $\mathbf{Q+k}_{HS}$. As we approach criticality, the physics is dominated by the hot spots and hence we should consider only the dispersion around these points. The rest of the FS is ``cold'' and can be disregarded. 
As we expand near the hot spots, we assume $|\delta \mathbf{k}|,| \mathbf{q}| \ll \Lambda$, where $\Lambda$ is the momentum cut off (of the order of $k_F$). The bare Green's function is therefore 
\begin{equation}
   G_{0}\left(\delta\mathbf{k+Q+k}_{HS},i\omega+i\Omega\right)=\left(i\omega-\mathbf{v}_{F,\mathbf{Q+k}_{HS}} \cdot \left(\delta\mathbf{k} +\mathbf{q}\right)\right)^{-1} , 
\end{equation}
 and the full Green's function is
\begin{align}
  &G \left(\delta\mathbf{k+k}_{HS},i\omega \right)=\\ \quad \notag
  &\left(i\omega-\mathbf{v}_{F,\mathbf{{k}_{HS}}} \cdot \delta\mathbf{k}-\Sigma \left( \delta\mathbf{k+k}_{HS},i\omega \right) \right)^{-1} . 
\end{align}
It can be shown that from symmetry considerations 
$\left|\mathbf{v}_{F,\mathbf{{k}_{HS}}} \right| =\left|\mathbf{v}_{F,\mathbf{Q+k}_{HS}} \right| \overset{\Delta}{=} v_F$.
As mentioned above, the angle between $\mathbf{v}_{F,\mathbf{{k}_{HS}}}$ and $\mathbf{v}_{F,\mathbf{Q+k}_{HS}}$ is $\theta$.

We compute $\Im\Sigma(\mathbf{k},\omega)$ From Eq.\eqref{Eq: Sigma_Matsubara} in the disordered phase ($1/\xi>0$) and at criticality ($1/\xi \to 0$). 
 We then obtain the spectral function by using the Kramers-Kroning relations. A detailed derivation of our analytical results is presented in Appendix \ref{appendix:analytical}. At the hot spot and in the disordered phase, we find 
\begin{align}
    \label{Eq:Im_Sigma_Exact1}
& \Im\Sigma(\mathbf{{k}_{HS}},\omega>0) \approx  \\ \notag
&    \begin{cases}
&  0 ,   \text{if} \ 0<\omega<\ v_{s}/\xi \\
&       -\lambda \ln \left[ \frac{v_{s}\sqrt{\omega^2 +\left(v_F /\xi\right)^2}}{v_F \left( \omega-\sqrt{\omega^2 -\left(v_{s}/\xi\right)^2} \right)}  \right],
            \ \text{if} \ \ v_{s}/\xi<\omega<v_s\Tilde{\Lambda} \\ 
      \\
&  -\lambda \ln \left[ \frac{\sqrt{\omega^2 +\left(v_F /\xi\right)^2}}{-\Lambda v_F+\sqrt{\omega^2 +\left(v_F \Tilde{\Lambda}\right)^2} }  \right],
      \  \text{if}\ \  v_{s} \Tilde{\Lambda} <\omega< E_F .\\
    \end{cases}
\end{align}
 Here $\Tilde{\Lambda} = \sqrt{\Lambda^2 +\xi^{-2}}$,  
 where in the vicinity of criticality, $\Tilde{\Lambda} \approx \Lambda$. 
 In addition, we have defined $E_F=v_F \Lambda$. As we have mentiond above,
Eq.\eqref{Eq:Im_Sigma_Exact1} is  valid only for $\omega \gg \frac{v_F}{v_S}\Omega_B$, where the undamped approximation is justified. However, for systems not too close to the QCP such that $\Omega_B \ll v_{s}/\xi$, the Landau damping term can be neglected even for $\omega \ll \Omega_B$. In this case, $D^{-1}\left(\mathbf{q},i\Omega_n\right)$ is dominated by the $\Omega^2/v_s^2$ term at high frequencies, and by the static term $\xi^{-2}$ at low frequencies. Assuming in addition that $1/\xi \ll \Lambda$, We find that
\begin{align}
\left. \frac{\partial \Re\Sigma(\mathbf{{k}_{HS}},\omega)}{\partial \omega}\right\vert_{\omega=0} =& \frac{1}{\pi} \int_{\infty}^{\infty}\frac{\Im\Sigma\left(\mathbf{{k}_{HS}},\Tilde{\omega}\right)}{\Tilde{\omega}^{^2}} \textrm{d}\Tilde{\omega}\approx  -\frac{\lambda \xi}{v_{s}}. 
\end{align}
This implies that the spectral function at the hot spot contains a quasi-particle piece at $\omega=0$, $A(\mathbf{{k}_{HS}},\omega) \sim Z \delta\left(\omega \right)$, with a quasi-particle weight $Z=\frac{v_{s}}{\xi\lambda}$ that vanishes as $1/\xi$ when approaching the QCP.
Interestingly, the opposite (overdamped boson) limit, $v_{s}\gg v_F$, gives the same parametric dependence of $Z$ on $\xi$ (see Appendix~\ref{appendox:scaling_Arguments}). .

At criticality and for higher frequencies, such that $\frac{v_F}{v_s}\Omega_B, v_s \Lambda \ll \omega \ll E_F$ , a simple analytical expression can be derived  for $A(\mathbf{{k}_{HS}},\omega)$ \begin{align}
A(\mathbf{{k}_{HS}},\omega)\approx -\pi^{-1} \Im \frac{1}{\omega+\frac{\pi}{2}\lambda \sign \left(\omega \right)+i\lambda \ln \left(\frac{E_F}{|\omega |}\right)} .
\label{Eq:A_QCP_HS}
\end{align} 
For $\lambda \ll E_F$,  $A(\mathbf{{k}_{HS}},\omega)$ has a local maximum (see Appendix~\ref{appendix:local_max_A_HS}) at   
\begin{equation}
\omega_{\rm{max}}=\lambda \left( \sqrt{ \frac{\ln{\frac{E_F}{\lambda} }}{2} } -\frac{\pi}{4} \right) +\mathcal{O}\left(\frac{1}{\sqrt{\ln{\frac{E_F}{\lambda}}}}\right) .    
\end{equation}
For consistency, the existence of such a local maximum requires that $v_s \Lambda < \omega_{\rm{max}}\sim \lambda$, up to logarithms. Notice that, for not too large $E_F/\lambda$, $\omega_{\rm{max}}$ is at the border of the underdamped regime (recall that $\lambda\sim \Omega_B v_F/v_s$). To check whether a local maximum obtains for reasonable values of $\lambda/E_F$, we need to numerically evaluate the spectral function without neglecting the damping term.

\begin{figure}
        	\centering\includegraphics[ width=1\columnwidth]{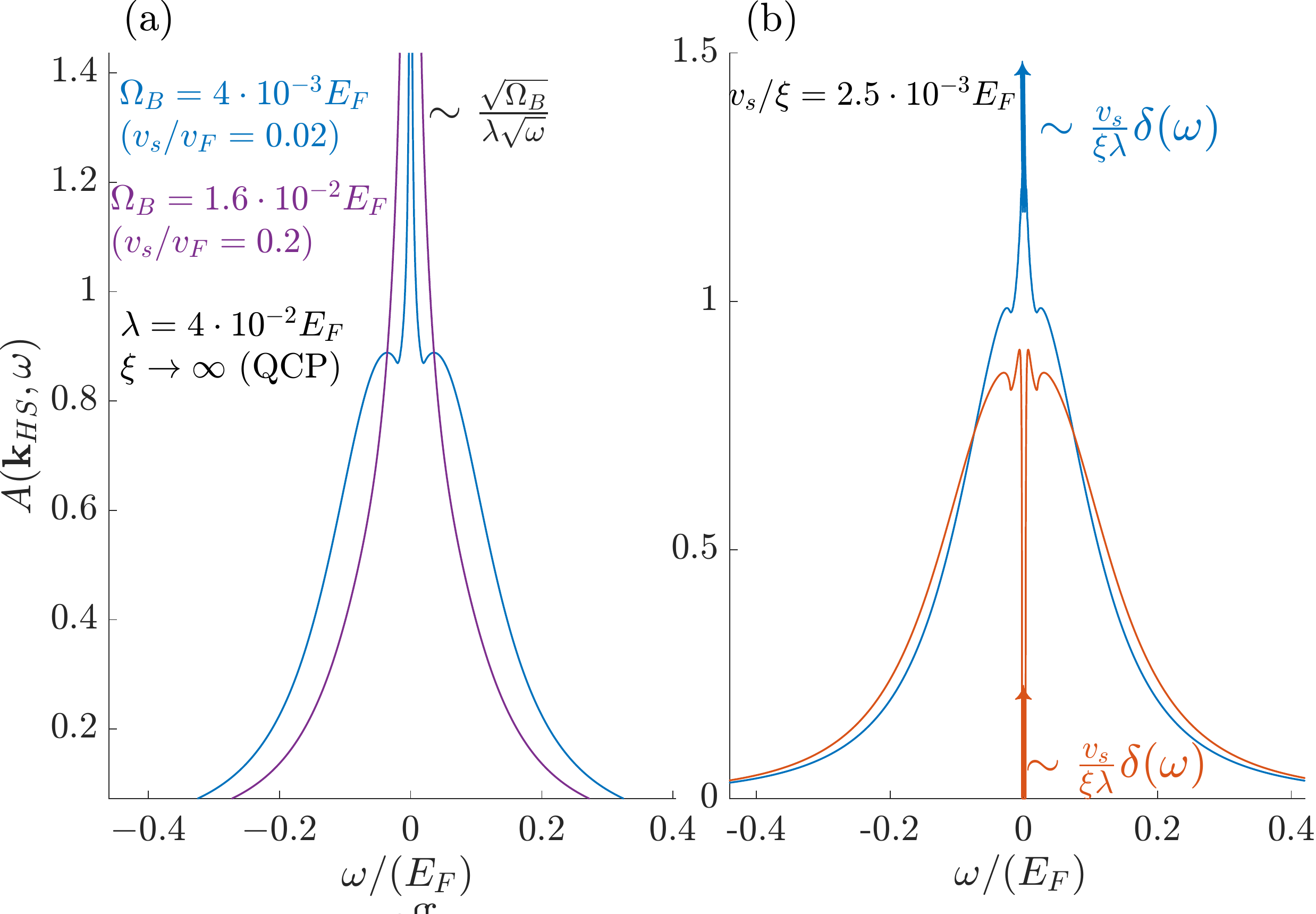}
	\caption{The spectral function at the hot spot, with damping (blue and purple) and without damping (orange).  (\textbf{a}) The behaviour at the critical point ($\xi \to \infty$). For sufficiently $\omega \ll \Omega_B$ we reproduce the known results of $\sim 1/\sqrt{\omega}$  divergence when finite damping is included (purple and blue). For sufficiently small $v_s/v_F \approx 0.02$  we see a local maximum behaviour  at $\omega_{\rm{max}}\sim\lambda$ (blue). (\textbf{b})  The behaviour with a finite correlation length $\xi$, with damping (blue) and without damping  ($\gamma = 0$, orange). The quasi-particle weight vanishes as $1/\xi$ in both cases.}
\label{fig:hot_spot}
\end{figure}

The overdamped regime at criticality ($\omega\ll \Omega_B$) has been well studied, and the spectral function is of the form $A \left( k_{HS},\omega \right) \sim \sqrt{\Omega_B}/\left( \lambda \sqrt{\omega} \right) $\cite{Sachdev2011}. 
To illustrate the behavior of the self energy at higher frequency, we present the full $\Sigma(\mathbf{{k}_{HS}},\omega)$ (evaluated numerically, see Appendix \ref{appendix:analytical}) in Fig.~\ref{fig:Self_Energy_Hs}. In the underdamped regime, $\omega\gg v_F\Omega_B/v_s$, $\mathrm{Im} \Sigma(\omega)\approx \lambda \mathrm{ln}(2E_F/|\omega|)$, as expected from Eq.~\eqref{Eq:Im_Sigma_Exact1}. 
The corresponding spectral function at the hot spot is shown in Fig.~\ref{fig:hot_spot}\textcolor{blue}{a} at criticality, for different values of $v_s/v_F$. For $v_s/v_F\approx 0.02$, the spectral function exhibits a local maximum at finite frequency -- a ``precursor gap'' feature. However, the local maximum disappears for $v_s/v_F=0.2$. 
At lower frequency, $\omega \lesssim \Omega_B \sim v_s \lambda/v_F$, the spectral function always diverges as $1/\sqrt{\omega}$. A similar plot in the away from the QCP ($1/\xi>0$, i.e., in the disordered phase) is shown in Fig.~\ref{fig:hot_spot}\textcolor{blue}{b}. In this case, a delta function quasi-particle peak appears at zero frequency, with a quasi-particle weight of $v_s/(\lambda \xi)$. 

\begin{figure}
        	\centering\includegraphics[ width=1\columnwidth]{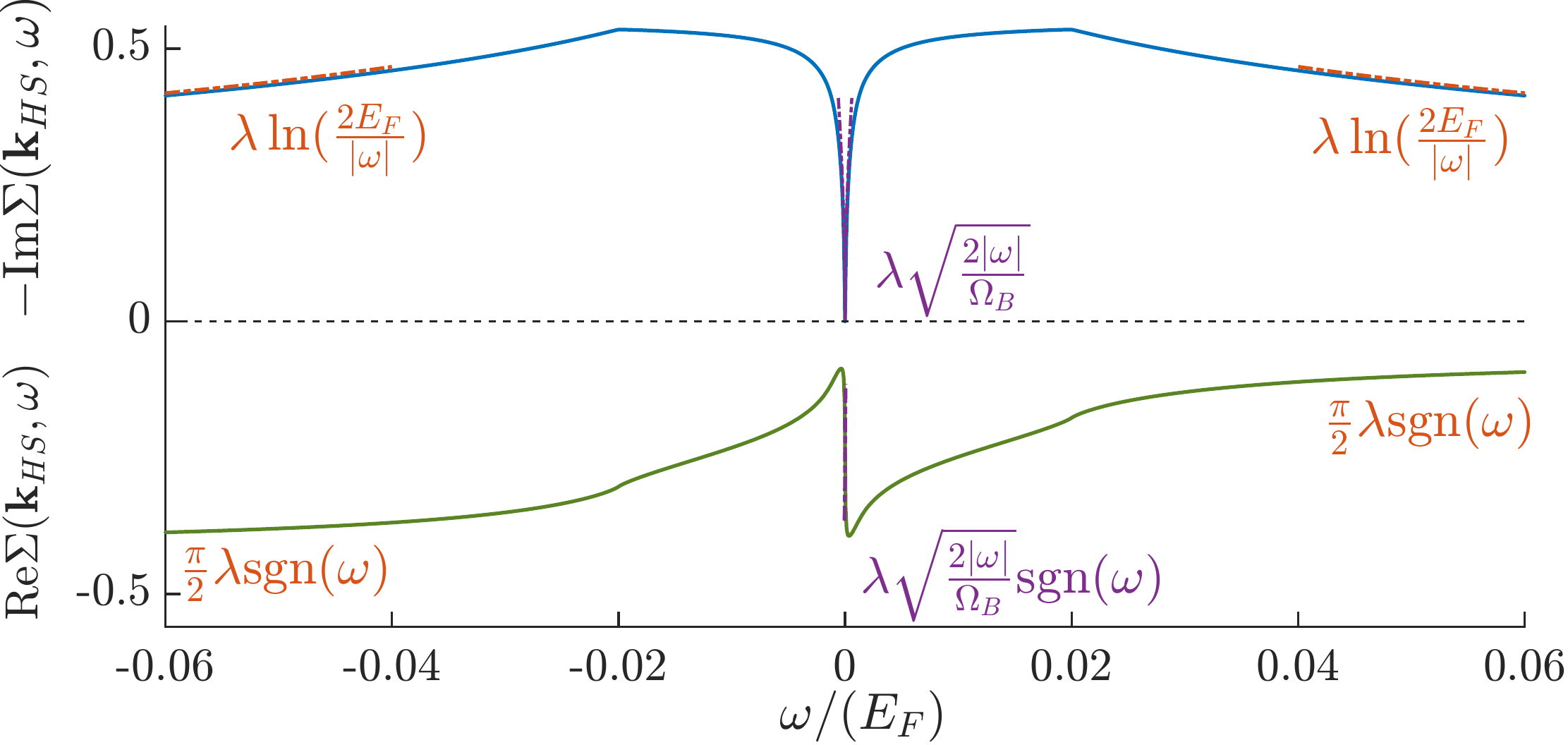}
	\caption{The self energy at the hot spot for $\xi\to\infty$. We mark the well known $\sqrt{\omega}$ singularity in the overdamped regime, and the log behaviour we have found in the undamped regime. Here $\lambda=4\cdot 10^{-3}E_F ,\Omega_B=1.6\cdot 10^{-3}E_F$ corresponding to $v_s/v_F=0.02$.}
\label{fig:Self_Energy_Hs}
\end{figure} 

\section{Numerical results in the quenched approximation}
\label{sec:numerical}
The results reviewed above were obtained from leading order perturbation theory. There is no obvious reason to trust these results all the way to the smallest frequencies and temperatures, 
especially near criticality ($\xi \to \infty$). In order to substantiate our conclusions, we have simulated the undamped model and calculated the Green's function in imaginary time.  
The simulation was done within the quenched approximation, in which we neglect fermionic corrections to the bosonic propagator~\cite{Meszena2016}. Within this approximation, a space-time configuration of the bosonic field $\vec{\phi}(\mathbf{r},\tau)$ is drawn at random from a Gaussian distribution given by $e^{-S_\phi}$. The fermion Green's function is calculated numerically for each such configuration. The results are then averaged over many configurations (see Appendix~\ref{appendix:simulation_process_GF}).   
    
    It is important to note that the class of diagrams summed within the quench approximation is larger than the standard ``rainbow diagrams'' (which are not sufficient in this case, see Appendix~\ref{appendix:Eliashberg}) . We illustrate some of the relevant diagrams in Fig. \ref{fig:diagram_quench}.
    The quenched approximation diagrams can be summed exactly if the AFM fluctuations are treated as static (only the $\Omega_n=0$ component is kept in the bosonic propagator)~\cite{Schmalian1999}. However, here, we are interested in the low-temperature limit, where this approximation is not justified. 
    We also note that in the case of $\mathbf{Q}=0$ ordering, e.g. at a ferromagnetic or nematic QCP, the fermionic self energy can be calculated analytically within the quenched approximation~\cite{Meszena2016}. This is because in that case, the problem becomes effectively one-dimensional upon linearizing the FS. This method does not generalize easily to our case, since the FS meet at the hot spot at a non-zero angle, and hence the problem is inherently two-dimensional.

In our simulations, we have used a lattice version of Eq.(\ref{Eq:model}) with a tight binding  dispersion, given by $
\epsilon_{\mathbf{k}}=2t_{A} \left[ \cos (k_x)+\cos (k_y) \right]   
+4t_{B} \cos(k_x)\cos(k_y)-\mu$ (we set the units such that the lattice spacing is equal to unity). In $S_\phi$, $\mathbf{q}^2$ is replaced by its lattice form: $4\left[\sin^2(q_x/2) + \sin^2(q_y/2)\right]$. 
We have used the following values for the microscopic parameters: $t_A=-0.85$ for the nearest neighbour (NN) hopping, $t_B=-0.45$ for the next nearest neighbour (NNN) hopping, and $\mu=1$. The purpose of including also NNN hopping is to avoid perfect nesting by $\mathbf{Q}=(\pi,\pi)$. We also set $g=1.5$ for the interaction strength, and $v_s=0.1$. At the hot spot, we find that for the above parameters, $v_F \approx 2$, and therefore $\lambda=(3g^2 v_{s})/(4\pi v_F) \approx 0.03$ and $v_s/v_F\approx 0.05$.  We carry out most of our simulations at $T=0.002$. This temperature is sufficiently small such that the results are representative of the behavior at $T=0$.

In the simulations, the imaginary time axis is discretized with a spacing of $\Delta\tau=0.5$. We have checked that the resulting Trotter error in the fermion self-energy is small for $\omega_n \lesssim \mu$. In  Appendix~\ref{appendix:size_effects}, We present further details regarding the simulation, and provide an analysis of the Trotter errors and finite size effects in both the spatial and Euclidean time axes.

\begin{figure}[t]
\centering\includegraphics[ width=1\columnwidth]{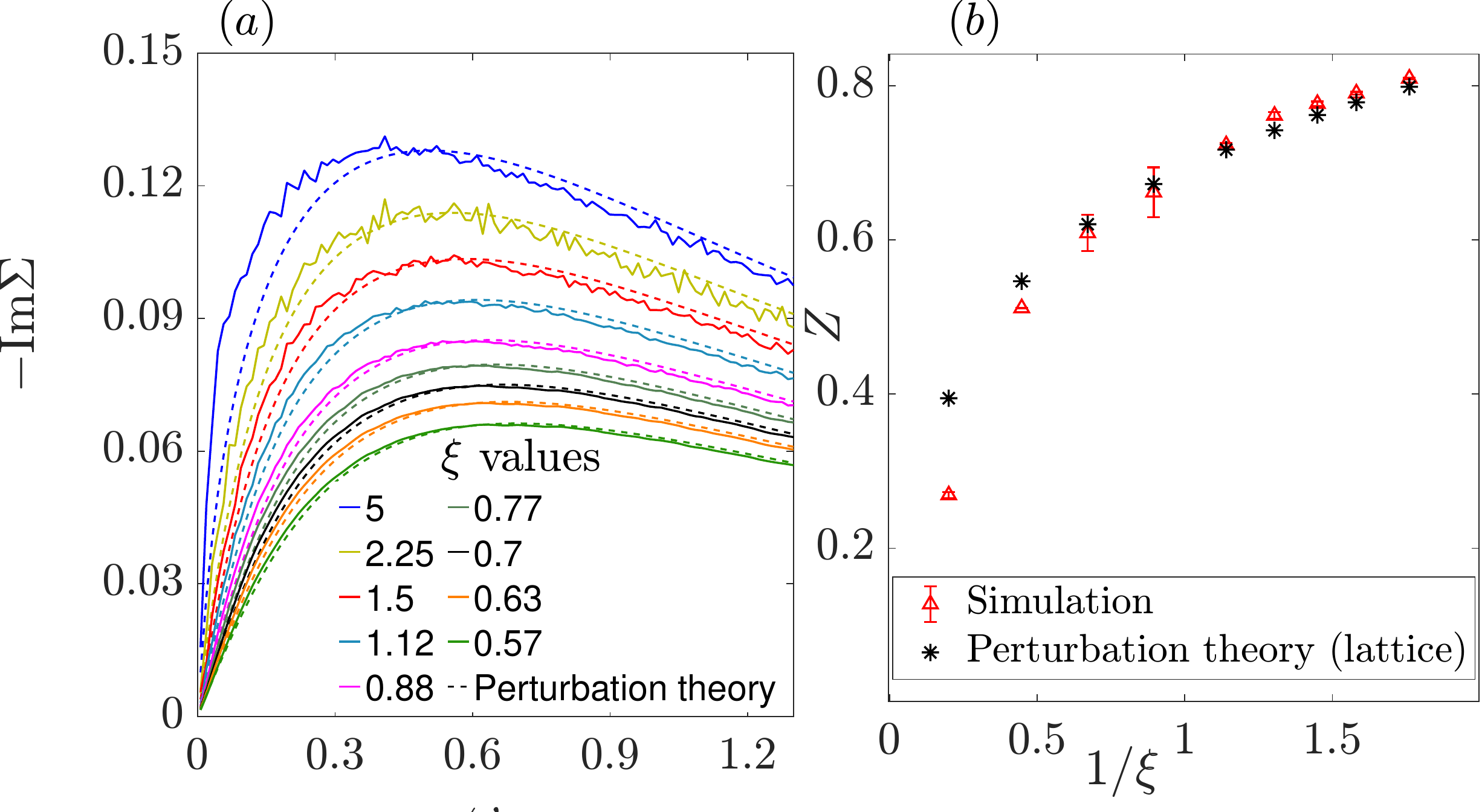}
	\caption{A comparison of the lattice model results, between the undamped perturbation theory and the quenched approximation simulation (for $L=16$). (\textbf{a}) The results for $\Im\Sigma(i\omega_n,\mathbf{k} \approx \mathbf{{k}_{HS}})$, for various $\xi$ values, from perturbation theory (dashed) and from the simulations of the quenched approximation (solid). (\textbf{b}) The results for the quasi-particle weight, $Z$, at $\omega=0$   from perturbation theory and from quenched approximation simulation}.
\label{fig: simulation_Sigma}
\end{figure}

In Fig~\ref{fig: simulation_Sigma}\textcolor{blue}{a} we compare the numerically exact $\Im\Sigma\left(i\omega_n,\mathbf{k \approx k_{HS}} \right)$ to the self-energy obtained from leading-order lattice perturbation theory for different values of $\xi$. 
We find an overall qualitative match between the two calculations, even at low frequencies such that $\omega_n \lesssim \Im\Sigma(i\omega_n,k \approx k_{HS})$, beyond the formal range of applicability of perturbation theory. 
The deviation between the perturbative results and the exact ones becomes substantial only at the largest value of $\xi$ and the lowest frequencies, where the exact self-energy is significantly larger than the perturbative one. 

We find a linear behavior of $\rm{Im}\Sigma(i\omega,\mathbf{{k}_{HS}})$ at the lowest Matsubara frequencies. From the slope at $\omega_n\rightarrow 0$ we can estimate the quasi-particle weight, $Z$, according to $Z=[1-\partial_\omega     \Re\Sigma(i\omega_n=0)]^{-1}$ (see Appendix~\ref{appendix:simulation_process_GF}). We summarize our results for $Z(\mathbf{k_{HS}})$  in Fig~\ref{fig: simulation_Sigma}\textcolor{blue}{b}.
The simulation results are very close to the perturbative ones for $\xi<2a$ ($Z \geq 0.6$). Closer to the critical point, the exact quasi-particle weight deviates downward from the perturbative one.

In addition, in order to verify that the behaviour of the self energy in our simulations is not a result of a nearby superconducting transition, we have measured numerically the superconducting susceptibility. According to the Berezinski-Kosterlitz-Thouless theory for a finite temperature XY transition, when plotting $\chi_{d,s} L^{-1.75}$  as a function of $T$ (where $L$ is linear system size), the intersection between the results for different system sizes should occur at $T_c$\cite{Moreo1991}.
However no indication for this crossing was found, and there is no signature of a rapid divergence which typically indicates the presence of superconducting fluctuations.  We therefore conclude that the observation of $Z \sim 1/\xi$ may be attributed to AFM fluctuations, as the perturbative calculation suggests (see our results in Appendix~\ref{appendix:simulation_process_sc}).

\section{Discussion}
\label{sec:discussion}
In this work, we have studied the electronic spectral function in the vicinity of an AFM metallic quantum critical point at $T=0$. We have found that, if the characteristic speed of the critical AFM fluctuations is significantly smaller than that of the electronic quasiparticles, the spectral function at the hot spot displays a local maximum at finite frequency -- a feature that can be interpreted as a precursor to the antiferromagnetic gap in the ordered phase. Away from the critical point, a finite quasiparticle peak, whose weight is inversely proportional to the AFM correlation length, remains at $\omega=0$. 

Our results can be understood qualitatively in terms of a simple physical picture. In the limit $v_s\ll v_F$, electrons whose energy $E$ is larger than the typical AFM fluctuation frequency (of the order of $v_s k_F$) experience an essentially static, spatially varying AFM order. If the magnetic correlation length is larger than $v_F/E$, we expect the electron spectral function to resemble that of metal with static, uniform AFM order. This can explain the appearance of an AFM precursor gap at the hot spots. Note that within our theory, the system is an ordinary FL obeying Luttinger's theorem all the way to the quantum critical point; hence, at zero frequency there is a quasi-particle peak at the hot spot, with a small but non-zero weight.

However, we find that observing a precursor gap at the hot spots requires a very small value of $v_s/v_F$, of the order of $10^{-2}$. In the cuprate superconductors, $v_s$ can be estimated from neutron and x-ray scattering measurements in the magnetically ordered state, whereas $v_F$ can be measured by angle-resolved photoemission. These estimates indicate that $v_F/v_s \approx 3$\cite{Lee2014,Bourges1997}. It therefore seems unlikely that simple antiferromagnetic fluctuations can explain the precursor gap feature observed in Ref.~\cite{He2019}.

\section{Acknowledgments}
We are grateful to A. Chubukov, S. Kivelson and Jörg Schmalian 
for illuminating discussions. 
This work was supported by the European Research Council (ERC) under the European Union’s Horizon 2020 research and innovation programme (grant agreement No 817799), the Israel-USA Binational Science Foundation (BSF), and the ISF Quantum Science and Technology grant no. 2074/19.

\bibliography{main_sdw.bbl}
\clearpage
\appendix
\onecolumngrid

\section{Analytical calculation - 
perturbation theory} \label{appendix:analytical}
\subsection{The full one-loop calculation}

We begin with the calculation of the self energy  to leading order, within the linearized model (Eq.\eqref{Eq: Sigma_Matsubara}).
As mentioned in the text, $\left|\mathbf{v}_{F,\mathbf{{k}_{HS}}} \right| =\left|\mathbf{v}_{F,\mathbf{Q+k}_{HS}} \right| \overset{\Delta}{=} v_F$, 
and the one-loop integral takes the form
\begin{align}\label{Eq: full_one_loop_int}
\Sigma \left( \mathbf{k+k}_{HS},i\omega \right)&= 3g^2\int_{q,\Omega} G_{0}\left(\mathbf{ q+Q+k+  k}_{HS},i\omega+i\Omega\right)D\left(-\mathbf{q},-i\Omega\right)\\ \quad \notag
&=3g^{2}\int_{q_x,q_y,\Omega} \frac{1}{i(\omega+\Omega)-v_{F}(k_{x}+q_{x})}\frac{1}{q^{2}+\gamma \left|{\Omega}\right|+\Omega^{2}/v_{s}^{2}+ \xi^{-2} }
\end{align}
where the factor 3 comes from the trace over different Pauli matrices. In addition, we are expanding around the hot spots, $\epsilon_{k+q} \approx v_{F} \left(k_{x}+q_x \right)$ as $|\mathbf{k}|,|\mathbf{q}|\ll 
\Lambda$ . In the following derivation we have chosen for simplicity $\mathbf{v}_{F,\mathbf{Q+k}_{HS}} || \hat{x}$. 
We start our discussion with the full one loop calculation. When $\gamma\neq 0$, we cannot directly integrate out $\Omega$ . In order to get around the non anlaticity of $D(\mathbf{q},i\Omega)$ we use the bosonic spectral function representation  
$D(\mathbf{q},i\Omega)= \int \frac{d\Omega'}{\pi} \frac{\mathrm{Im}D(\mathbf{q},\Omega')}{i\Omega -\Omega'}$
where $D(\mathbf{q},i\Omega')=\left( q^2 +\xi^{-2}+\Omega'^2/v_{s}^2 +\gamma|\Omega'| \right)^{-1}$. A straightforward analytical continuation (on the upper half plain, $i\Omega' \to \Omega'+i\eta$)  gives $\mathrm{Im}D(\mathbf{q},\Omega')=\left(\gamma\Omega'\right)/ \left( \left( \xi^{-2}+q^{2} -\Omega'^2 /v_{s}^2 \right)^{2} +\gamma^2\Omega'^2 \right)$. Plugging this in the one loop integral yields
\begin{align}
\label{Eq:damped_self_energy}
\Sigma \left( \mathbf{k+k}_{HS},i\omega \right)& =6g^{2}\int_{q_x,q_y,\Omega',\Omega} \frac{1}{i(\omega+\Omega)-v_{F}(k_{x}+q_{x})} 
\frac{1}{i\Omega-\Omega'}
\frac{\gamma\Omega'}{ \left( \xi^{-2}+q^{2} -\Omega'^2 /v_{s}^2 \right)^{2} +\gamma^2\Omega'^2 } .
\end{align}
We  can now perform the integral $d\Omega$ using Cauchy's theorem and analytically continue  to the upper half plain ($i\omega \to \omega+i\eta$). Doing so, together with applying Dirac identity for the imaginary part, we  are left with the following integral
\begin{align}
\label{Eq:theta_functions}
\Im\Sigma(k,\omega) &= 6\pi g^2 \int_{q_x,q_y,\Omega'} D(\Omega', \mathbf{q}) \left[ \Theta \left(-\epsilon_{k+q} \right)  - \Theta \left(-\Omega' \right) \right] \delta \left(-\epsilon_{k+q} +\omega+\Omega' \right)  \\ \quad \notag
    &=  3 g^2 \int_{q_x,q_y} D(\epsilon_{k+q}-\omega, \mathbf{q}) \left[ \Theta \left(-\epsilon_{k+q} \right)  - \Theta \left(\omega-\epsilon_{k+q} \right) \right]  \\ \quad \notag
&=
3g^{2}\int^{\Lambda}_{-\Lambda}\frac{dq_{y}}{4\pi^{2}}\int_{-k_x}^{\tilde{\omega}/v_{F}}dq_{x}\frac{\gamma\left(\tilde{\omega}-v_{F}q_{x}\right)}{\left[q_{x}^{2}+q_{y}^{2}-\frac{\left(\tilde{\omega}-v_{F}q_{x}\right)^{2}}{v_{s}^{2}}\right]^{2}+\gamma^{2}\left(\tilde{\omega}-v_{F}q_{x}\right)^{2}}\\ \quad \notag
&=
  -3g^{2}\Im\int^{\Lambda}_{-\Lambda}\frac{dq_{y}}{4\pi^{2}}\int_{-k_x}^{\tilde{\omega}/v_{F}}dq_{x}\frac{1}{\left[q_{x}^{2}+q_{y}^{2}-\frac{\left(\tilde{\omega}-v_{F}q_{x}\right)^{2}}{v_s^{2}}\right]+i\gamma\left(\tilde{\omega}-v_{F}q_{x}\right)} . \\ \quad \notag 
\
\end{align}
From now on, we will focus on studying the behaviour at the hot spot ($\mathbf{k}=0$).We perform the  $q_{y}$  integral analytically (the  remaining integral over $q_{x}$ is also doable but too complicated be useful). We find that
\begin{align}
    \Im\Sigma(0,\omega)&=
    \frac{3g^2 c}{4\pi^2}
    \int_{0}^{\omega/v_{F}}dq_{x}{\rm Im}\left(\frac{\tan^{-1}\left(\frac{v_s\Lambda}{\sqrt{-(v_{F}q_{x}-\omega)^{2}+v_s^{2}\left(\xi^{-2}+q_{x}^{2}-i\gamma\omega+i\gamma v_{F}q_{x}\right)}}\right)}{\sqrt{-(v_{F}q_{x}-\omega)^{2}+v_s^{2}\left(\xi^{-2}+q_{x}^{2}-i\gamma\omega+i\gamma v_{F}q_{x}\right)}}\right) .
\end{align}
Note that for $\omega\approx v_s\Lambda$ and small damping constant, such that $v_s\gamma\ll \Lambda$, the numerator takes the form of  
$\tan^{-1}({\sqrt{v_s\Lambda}}/{\sqrt{-v_s\Lambda +2v_F q_x}})$ in the vicinity of $q_x=0$.
Crossing $q_x=0$ leads to discontinuity that originates from the jump in phase when the sign of the $\tan^{-1}$ argument is reversed. This leads to a cusp behaviour at $\omega \approx v_s\Lambda$.

At the small $\omega$ regime ($\omega\ll \Lambda^2/\gamma,v_s\Lambda $) and close to criticality ($ \Lambda\xi \gg  1$) it can be shown that the integral over $q_x$ boils to
\begin{align}
\Im\Sigma(0,\omega) &= \frac{3g^2}{4\pi} \Im \int^{\omega}_{0} \frac{d q_x}{\sqrt{D-Ci}}
\end{align}
with   $C\equiv\gamma \left(v_{F}q_{x}-\omega\right)$,$D\equiv\xi^{-2}+q_{x}^2 -[C/(v_s\gamma)]^2 $.
This $dq_x$ integral is doable, and we get
\begin{align}
 \Im\Sigma(\omega,k) =  \left. \frac{3g^2}{4\pi} \Im \left\{ \frac{1}{\sqrt{1-v_F^2/v_{s}^2}}\ln\left[2\left(q+\sqrt{\beta+q\left(\alpha +q \right)}\right)+\alpha\right] \right\} \right \vert^{q=\omega/v_{F}}_{q=0}
 \label{Eq:Im_Sigma_general}
 \end{align}
where $\alpha \equiv \frac{2\omega v_{F}-i\gamma v_{F} v_{s}^2 }{v_{s}^2-v_F^2} $ and $\beta \equiv \frac{v_{s}^2 \left(i\omega\gamma +\xi^{-2}\right) -\omega^2}{v_{s}^2 -v_{F}^2}$. 
Expanding for small $\omega$, we find the standard quadratic term of a FL
 \begin{align}
 \Im\Sigma(0,\omega) \approx -\frac{3\gamma g^2\xi^{3} \omega^2}{8\pi v_{F}}
\label{Eq:Im_Sigma_damped}
\end{align}

Expanding  for small $\omega$ at the QCP ($\xi\to\infty$), we find the well known form of the self energy\cite{Sachdev2011}
\begin{align}
   \Im\Sigma(0,\omega) =-\lambda\sqrt{\frac{2\abs{\omega}}{\Omega_B}}
\end{align}
with $\lambda \equiv\frac{3g^2}{4\pi v_F}$. Since this is a power law behaviour we can find directly the Matsubara form and the real part. We conclude that
\begin{align}
\label{Eq:Matsubara_qcp}
    \Sigma(0,i\omega) &=2i\lambda\sqrt{\frac{\abs{\omega}}{\Omega_B}}\sign\left(\omega\right) \\
     \Sigma(0,\omega) &=-\lambda\sqrt{\frac{2\abs{\omega}}{\Omega_B}}\left(i+\sign(\omega)\right)  
\end{align}
 
Let us now study the crossover to the undamped regime. We start from the bottom line in Eq.~(\ref{Eq:theta_functions}).

We integrate over $q_x$ first, yielding

\begin{align}
  \Im\Sigma(0,\omega>0)=- \frac{3g^2c}{2\pi^2}\Im \int dq_y \left( \frac{\tan^{-1}\left(\frac{\frac{v_F}{v_s}(\omega-i\Omega_B)}{\sqrt{4\Omega_B\omega-4\omega^2+\left(\frac{v_F}{v_S}\Omega_B\right)^2 +4(v_s^2-v_F^2)q_y^2}}\right)}{\sqrt{\Omega_B\omega-\omega^2+\left(\frac{v_F}{v_S}\Omega_B\right)^2 +4(v_s^2-v_F^2)q_y^2}} \right)-
   \left( \frac{\tan^{-1}\left(\frac{\frac{v_s}{v_F}(\omega-i\left(\frac{v_F}{v_S}\right)^{2}\Omega_B)}{\sqrt{4\Omega_B\omega-4\omega^2+\left(\frac{v_F}{v_S}\Omega_B\right)^2 +4(v_s^2-v_F^2)q_y^2}}\right)}{\sqrt{\Omega_B\omega-\omega^2+\left(\frac{v_F}{v_S}\Omega_B\right)^2 +4(v_s^2-v_F^2)q_y^2}} \right) 
\end{align}
where $\Omega_B=v_s^2\gamma$ as in the main text. Note that we  can ignore the second term for $v_s\ll v_F$. 

Despite the complicated integrand, we can understand the behaviour at large $\omega$. First, we now clearly see that in order to be able to neglect all the $\gamma$-dependent terms, we must require
\begin{align}
\omega \gg \frac{v_F \Omega_B}{v_s} .
\end{align}
At this large $\omega$ regime and together with $v_F/v_s \gg 1$, we find that the integral becomes very simple
\begin{align}
\Im\Sigma(0,\omega>0) \approx -\frac{3g^2 c}{8\pi}
 \int^{\Lambda}_{\Lambda} \frac{dq_y}{\sqrt{\omega^2+ v_F^2q_y^2}} =-\lambda\ln(\frac{2E_F}{\omega})
\end{align}
yielding exactly the dependence we are expecting to have in the undamped regime (see Eq.\eqref{Eq:A_QCP_HS}).
\subsection{Undamped model: $\Im(\Sigma)$}
\label{appendix: udnamped_detailed}
We now focus on the limit in which we can neglect damping, taking $\gamma=0$. 
As mentioned in the main text, this assumption is not fully justified as the crossover to the undamped regime and the local maximum in the spectral function occur the same energy scale. As we approach criticality, we have no parametric separation between the two and we cannot describe the pseudo-gap behaviour analytically. However, if $\Omega_B \ll \lambda$, damping effects become less important, and at least qualitatively we find a similar behaviour to the undamped case (see Fig.~\ref{fig:Self_Energy_Hs}). Inspired by that, we turn  to study this case in more details. 

\begin{figure}
	\hspace{-0.18cm}
 	\centering\includegraphics[ width=1\columnwidth]{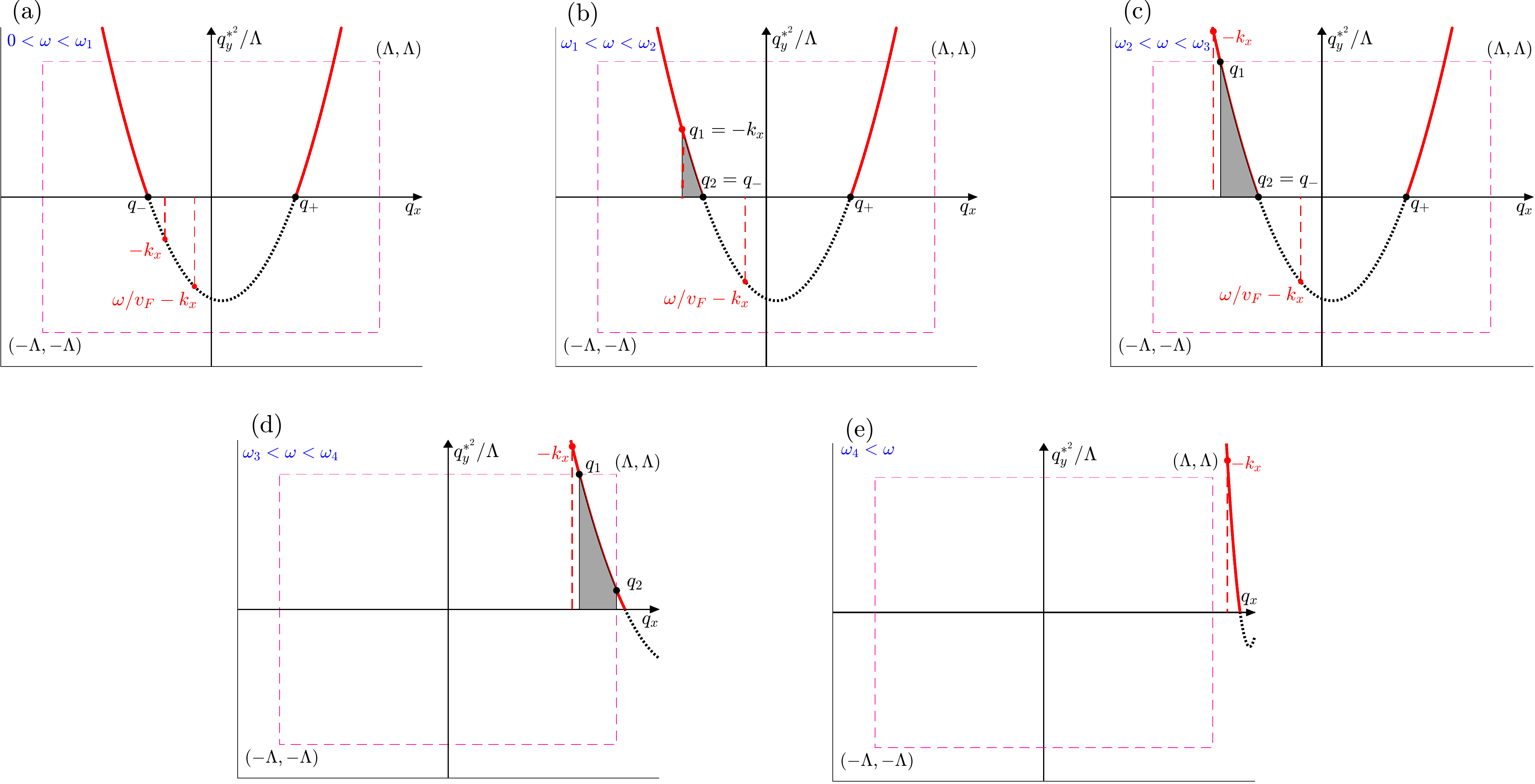}
	\caption{Panels (a-e) describe schematically the different integration limits $q_1,q_2$ in Eq.\eqref{Eq:integral_qx}. In red, we mark the constrains on $q_{y}^{*}$ from the $\delta$ function (dashed line and solid line for the conditions obtained from Eq.\eqref{Eq:delta_sol} and Eq.\eqref{Eq:qy_sol} respectively)  . In addition, the dashed pink square denotes the cutoff restriction. As we change $\omega,k$, we have a different $q_{y}^{*^{2}}$ , and the integration area is changed accordingly. The exact expressions for $\omega_{1,2,3.4}$ which separate between the regimes are mentioned in Eq.\eqref{Eq: omega_values} and can be deduced from the Intermediate cases between the different panels. }
\label{fig:Integral_limit}
\end{figure}

The one-loop integral takes the form
\begin{align}
\Sigma \left( \mathbf{k+k}_{HS},i\omega \right)& =3g^{2}\int_{q_x,q_y,\Omega} \frac{1}{i(\omega+\Omega)-v_{F}(k_{x}+q_{x})}\frac{1}{q^{2}+\Omega^{2}/v_{s}^{2}+ \xi^{-2} } .
\end{align}
Since damping is neglected, the Cauchy integration over $\Omega$ is straightforward. Once we perform the frequency integral we can perform an analytical continuation. We can then apply Dirac identity, and obtain the imaginary part of $\Sigma$ in real frequency
\begin{align}
\Im\Sigma(\mathbf{k},\omega) & = \frac{-3\pi g^2 v_{s}}{2} \int_{q_x,q_y}  \ \delta \left( \omega-v_{s} \cdot \sign \left(k_x+q_x \right) \sqrt{ q^2 +\xi^{-2}} -v_F \left(k_x +q_x \right)   \right) \frac{1}{\sqrt{q^2 +\xi^{-2}}}
\end{align}
where for simplicity of notation we have placed the origin at $\mathbf{{k}_{HS}}$, and therefore $\mathbf{k+k}_{HS} \rightarrow \mathbf{k}$ as before. We integrate over $q_y$ first. Note that it is enough to study the case $\omega>0$ since  the calculation is exactly the same for $k_x \to -k_x \And \omega \to -\omega $ from particle hole symmetry  (it can be seen by changing variables $q_{x,y} \to -q_{x,y}$ in the integral). We therefore focus on  the case $\omega>0$. Solving the $\delta$ function gives
\begin{align}
\frac{\Tilde{\omega}-v_{F}q_{x} }{\sign \left(k_{x}+q_{x} \right)}=v_s \sqrt{q^2 +\xi^{-2}}
\label{Eq:delta_sol}
\end{align}
where we have defined $\Tilde{\omega}\equiv\omega-v_{F}k_{x}$.
 Respecting the constraint of a positive LHS in Eq.\eqref{Eq:delta_sol}, given that $\omega>0$, leads to the requirement   $-k_{x}<q_{x}<-k_{x}+\omega/v_{F}$. One finds that the solution of the $\delta$ function for $q_y$ is
\begin{align}
q_{y}^{*} =\pm \sqrt{ \left( \frac{\Tilde{\omega}-v_{F}q_{x} }{v_{s}}
\right)^{2}-q_x ^2 -\xi^{-2} } .
\label{Eq:qy_sol}
\end{align}
In order to have a real solution for $q_{y}$, a positive argument for the square root in Eq.\eqref{Eq:qy_sol} is also needed. One finds  that after performing the $\delta$ function integral over $q_y$ we are left with
\begin{align}
\Im\Sigma(\mathbf{k},\omega) & = -\frac{3 g^2}{4\pi} \int^{q_2}_{q_1} \frac{d q_x}{|q_{y}^* |} .
\end{align}
In order to determine $q_1,q_2$ we carefully consider the conditions which guarantee a well-defined solution for $q_y^{*}$, as mentioned above. We should also bring into account the momentum cutoff off $\Lambda$.
Practically, we find the overlap of the following constrains:
(1)  $-k_{x}<q_{x}<-k_{x}+\omega/v_{F}$ , (2)  $(q^{*}_y)^{2}>0$  (3) $\left| q_{x,y} \right| <\Lambda$.
Combining all three conditions gives the suitable integration limits $q_1,q_2$ (for a given $k_x,\omega$). An illustration of the integration area in the different regimes of $\omega,k_x$ is presented in Fig.~\ref{fig:Integral_limit}.
For example, if $\omega<v_s\sqrt{k_{x}^2 +\xi^{-2}}$ one can see that for any $q_{x} \in \left[-k_x, -k_x +\omega /v_F \right]  $ there is no real solution for Eq.\eqref{Eq:qy_sol}. On the other hand, $\omega$ is bounded from above by the cutoff restriction  for $q_y$.      
For fixed $q_{1},q_{2}$ we  find that the integral yields

\begin{align}
\Im\Sigma(\mathbf{k},\omega) & = -\frac{3 g^2}{4\pi} \int^{q_2}_{q_x=q_1} \frac{d q_x}{|q_{y}^* |}=  \left. -\frac{3 g^2}{4\pi \sqrt{v_{F}^2-v_{s}^2}} \ln \left[ \left(2\sqrt{q_x^2+Aq_x+B}+x \right)+A \right] \right \vert_{q_x=q_1}^{q_x=q_2}  
\label{Eq:integral_qx}
\end{align}
where we define $A \equiv -\frac{2\Tilde{\omega}v_{F}}{v_{F}^2-v_{s}^2} , B \equiv \frac{\omega^2 -(v_{s}/\xi)^2 }{v_{F}^2-v_{s}^2} $. From here, one can obtain the imaginary part for any $\mathbf{k},\omega$. 
We summarize our final results under the  assumption ${v_s} \ll v_F$:
\begin{equation}
   \Im\Sigma(k_x,\omega>0) \approx
    \begin{cases}
     0 , &  \text{if} \ 0<\omega<\omega_1\\

           -\lambda \ln \left[\frac{v_s\sqrt{\Tilde{\omega}^2 +(v_F^2-v_{s}^2)\xi^{-2}}}{\Tilde{\omega}v_F+(v_F^2-v_{s}^2)k_x-\sqrt{v_F^2-v_{s}^2}\sqrt{\omega^2-(v_s k_x)^2-(v_{s}/\xi)^{2}}} \right] ,
           &  \text{if} \ \omega_1<\omega<\omega_2\\ 

      \\
      -\lambda \ln \left[\frac{\sqrt{\xi^{-2}+\frac{\Tilde{\omega}^2}{v_F^2-v_{s}^2}}}{-\Lambda+\sqrt{\xi^{-2}+\Lambda^2+\frac{\Tilde{\omega}^2}{v_F^2-v_{s}^2}}}    \right] ,
      \ & \text{if}\ \ \omega_2<\omega<\omega_3\\
    \\

      -\lambda \ln \left[ \frac{(v_F^2-v_{s}^2)\Lambda +\sqrt{v_F^2-v_{s}^2}\sqrt{\Tilde{\omega}-v \Lambda)^2-(c\Lambda)^2-v_{s}^2}-\Tilde{\omega}v_F}{-v_{s}\sqrt{\Tilde{\omega}^2+(v_F^2-(v_{s}/\xi)^2)(\xi^{-2}+\Lambda)^2}+v_{s} \sqrt{v_F^2-v_{s}^2}\Lambda}   \right] ,
          \ & \text{if}\ \ \omega_3<\omega<\omega_4\\
        \\
0
          \ & \text{if}\ \ \omega_4<\omega\\
    \end{cases}
    \label{Eq:Im_Sigma_Exact}
  \end{equation} 
  
where we have used the following definitions
   \begin{align}
    \begin{cases}
   \omega_1=v_s\sqrt{\xi^{-2}+k_{x}^{2}}\\
   \omega_2=v_s\sqrt{k_x^2+\Lambda^2+\xi^{-2}} \\
   \omega_3=  v_Fk_x+E_F+v_{s} \sqrt{\Lambda^2+\xi^{-2}} \\
   \omega_4= v_Fk_x+E_F+v_{s} \sqrt{2\Lambda^2+\xi^{-2}} \\
    \end{cases}
    \label{Eq: omega_values},
    \end{align} 

   and $E_F\equiv v_F\Lambda$ as in the text. Focusing on the most interesting case, at the hot spot ($k_x=0$), one can easily show that  we reproduce the analytical results from the main text (Eq.\eqref{Eq:Im_Sigma_Exact1}). In addition, for $\omega_2<\omega<\omega_3$ and  $\omega \ll E_F$  it can be shown that $\Im\Sigma(k_x,\omega>0) \approx -\lambda\ln(\frac{2E_F}{\omega})$ in agreement with the results for $\Im\Sigma(0,\omega)$ that one can read from  Eq.\eqref{Eq:A_QCP_HS}.
   
     \subsection{Undamped model
     -$\Re\Sigma$ and quasi-particle weight at $\omega=0$}
  \label{appendix:Undamped_Real_part}
  
  To study the spectral function, we need also the real part of $\Sigma(\mathbf{k},\omega)$. This was obtained numerically using the Kramers-Kroning relations. Nevertheless We can study $\Re \Sigma(\mathbf{k},\omega)$ close to the FS  and close to criticality ($1/\xi \ll \Lambda$ ) in the undamped case. We find the quasi particle weight according to 
\begin{align}
\left. \frac{\partial \Re\Sigma(0,\omega)}{\partial \omega}\right\vert_{\omega=0} =& \frac{1}{\pi} \int_{\infty}^{\infty}\frac{\Im\Sigma\left(0,\omega'\right)}{\omega'^{^2}} \textrm{d}\omega'\approx  -\frac{\lambda \xi}{v_{s}}. 
\end{align}
Where the assumption of $\xi^{-1} \ll \Lambda$ allows us to use only the regime $\omega_{1} < \omega < \omega_{2}$ where the imaginary part is the most singular when approaching to criticality (for $v_s/\xi \lesssim \omega \ll v_s\Lambda$, $\Im\Sigma(0,\omega) \sim -\lambda\sqrt{\frac{\omega-v_s/\xi}{v_s/\xi}}$).

  \subsection{Spectral function at the hot spot}
   \label{appendix:local_max_A_HS}
Let us focus on the hot spot results in Eq.
\ref{Eq:A_QCP_HS}. Finding the local maxima by equating the derivative to zero yields (for simplicity we look for $\omega>0$ solution as the function is even)
\begin{align}
    \lambda^2 \ln^{2}\left(\frac{E_F}{\omega}\right)=2\omega\left(\omega+\frac{\pi}{2}\lambda\right)\ln\left(\frac{E_F}{\omega}\right)+\left(\omega+\frac{\pi}{2}\lambda \right)^2
    \label{Eq:maxima_full_Eq}
\end{align}
As Eq.\eqref{Eq:A_QCP_HS} is valid only for $\omega 
\ll E_F$ , the logarithm must be a large number. If we neglect the second term in the RHS, we are left with the following equation 
\begin{align}
  \lambda^2 \ln\left(\frac{E_F}{\omega}\right)=2\omega\left(\omega+\frac{\pi}{2}\lambda\right)
\label{Eq:maxima_simple_Eq}
\end{align}
therefore $\omega_{\mathrm{max}} \sim\lambda\sqrt{\ln\left(\frac{E_F}{\lambda}\right)}$ and the assumption regarding Eq.\eqref{Eq:maxima_full_Eq} is indeed self consistent (i.e, all terms have power $\alpha=1.5$ or $\alpha=2$ in $\ln^{\alpha}\left(\frac{E_F}{\omega_{\mathrm{max}}}\right)$ except the the neglected one with $\alpha=1$). Solving the quadratic equation.~\ref{Eq:maxima_simple_Eq} we find
\begin{align}
    \omega=\frac{-\pi\lambda+\sqrt{\left(\pi\lambda\right)^2 +8\lambda^2\ln\left(\frac{E_F}{\omega}\right)}}{4} =\lambda\left(\sqrt{\frac{\ln\left(\frac{E_F}{\omega}\right)}{2}}-\frac{\pi}{4} +\mathcal{O}\left(\sqrt{\frac{1}{\ln\left(\frac{E_F}{\omega}\right)}} \right)\right)
    \label{Eq:iteration}
\end{align}
Where the last equality is nothing but applying the self consistent assumption that led to Eq.\eqref{Eq:maxima_simple_Eq} (including higher order terms, can generate powers of $\alpha \leq 1$ in Eq.\eqref{Eq:maxima_full_Eq} ). We solve Eq.\eqref{Eq:iteration} iteratively, staring from $\omega_1=\lambda$. The second step includes only logarithmic corrections. A straightforward substitution  yields
\begin{align}
\omega_2=\omega_{\mathrm{max}}=\lambda\left(\sqrt{\frac{\ln\left(\frac{E_F}{\lambda}\right)}{2}}-\frac{\pi}{4} +\mathcal{O}\left(\sqrt{\frac{1}{\ln\left(\frac{E_F}{\lambda}\right)}} \right)\right).
\end{align}

We can go further to $\omega_3$ with $
\ln\ln\left(\frac{E_F}{\lambda}\right)$ corrections. However, in this case we cannot ignore any more higher order corrections in the original equation (Eq.\eqref{Eq:maxima_full_Eq}).  Furthermore, we assume $\lambda \ll E_F$  and hence $\omega_2$ is already a good estimation for the maximum.

 \subsection{$\Re\Sigma$ with damping} 
 \label{appendox:scaling_Arguments}
Similar to Section~\ref{appendix:Undamped_Real_part} for $\gamma=0$, we can also find the quasi particle weight at finite $\gamma$ (to  leading order in $\xi^{-1}$).  As before, we use Kramers-Kroning integral for the most singular part. Here the singular part is at  $\omega=0$. Using Eq.\eqref{Eq:Im_Sigma_general} in the vicinity of $\omega=0$ (and then expanding to leading order in $1/\xi$), and following the same calculation we find  
\begin{align}
\left. \frac{\partial \Re\Sigma(0,\omega)}{\partial \omega}\right\vert_{\omega=0} =& \frac{1}{\pi} \int_{\infty}^{\infty}\frac{\Im\Sigma\left(0,\omega'\right)}{\omega'^{^2}} \textrm{d}\omega'\approx  -\frac{\lambda \xi}{v_{s}}. 
\label{Eq:Re_Sigma_damped}
\end{align}
 
    Although the perfect agreement between the undamped and overdamped cases seems surprising, it can be justified from scaling arguments. Considering Eq.\eqref{Eq: full_one_loop_int}, and rescaling everything with respect to $\omega$  in order to have a unit-less integral,  leads to 
    \begin{align}
        \Sigma(0,i\omega)=|\omega|^{\frac{1}{2}}f(\xi^2\omega)
    \end{align}
    for small enough $\omega$.
   The scaling function $f$ satisfies $f(x\to\infty) \to \mathrm{const.}$, in agreement with Eq~\ref{Eq:Matsubara_qcp}.  On the other hand, for the case $f(x\to 0)$ we have a FL and hence a linear behavior  in $\Re\Sigma$ and quadratic behaviour in $\Im\Sigma$, as a function of real $\omega$, is expected. When switching to Matsubara frequencies, this leads to $\Im f(x\to 0) \sim \sqrt{x} $ and  $\Re f(x\to 0) \sim x^{\frac{3}{2}} $.  This immediately yields the powers of $\xi^{3}$ and  $\xi$ obtained in Eq.(\ref{Eq:Im_Sigma_damped}, \ref{Eq:Re_Sigma_damped}) respectively. The factors of $v_F,\gamma$ can then be deduced by dimensional analysis.  

 \subsection{Failure of the rainbow summation for undamped bosons ($\gamma=0$)}
 \label{appendix:Eliashberg}
\begin{figure}
      	\centering\includegraphics[ width=0.4\columnwidth ]{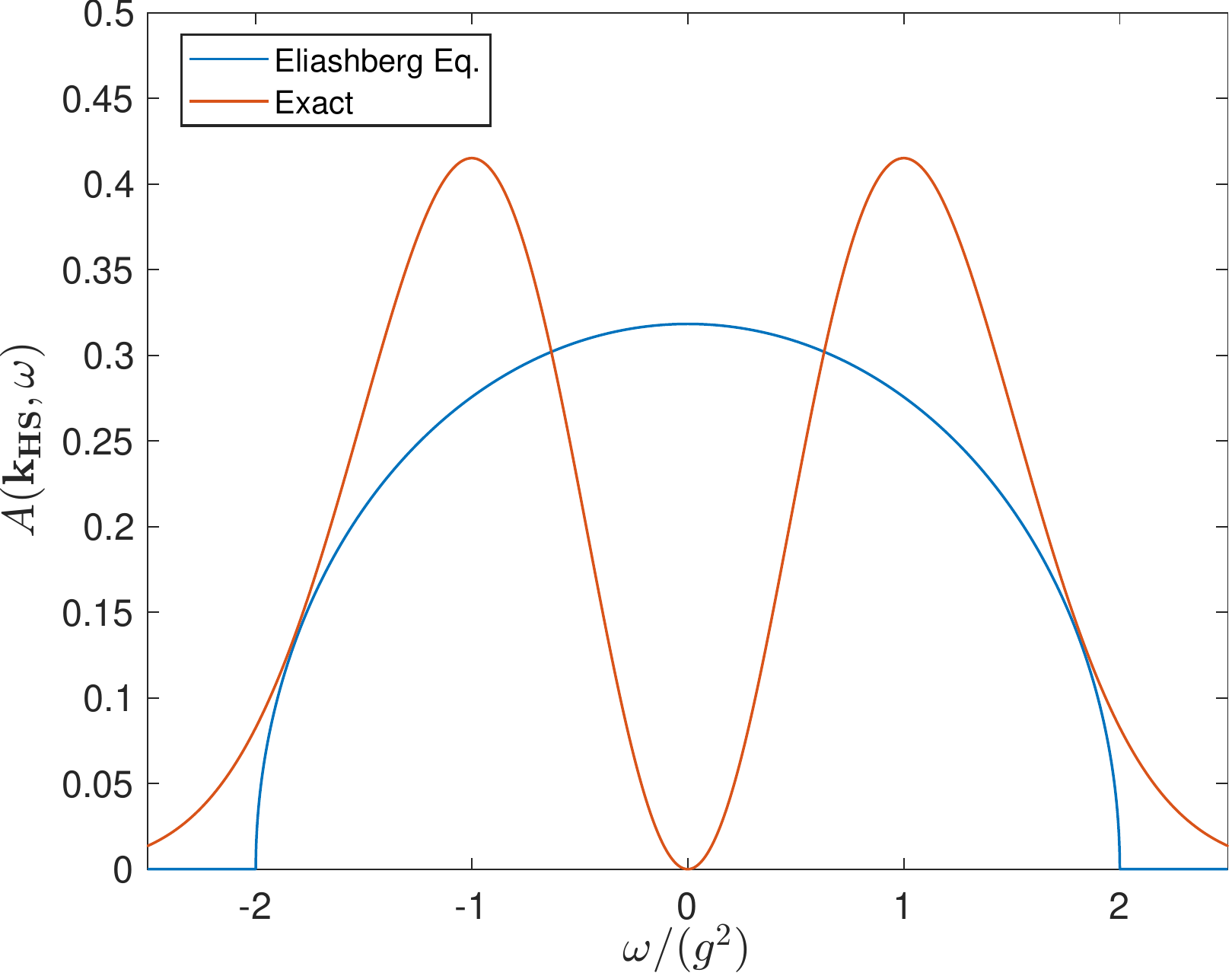}
	\caption{Comparison between two calculations of the spectral function, with a bosonic propagator : $D=\delta (\mathbf{q-Q},\omega)$ }
\label{fig: SC_p_Eliashberapg}
\end{figure}

Here, we demonstrate briefly how the `rainbow'' summation of self-energy diagrams fails close to criticality and predicts an incorrect spectral function. For simplicity, we study a case where the system is at the QCP ($\xi \rightarrow \infty$), but the bosonic propagator has no dynamics. 
Let us consider a toy model with the same $S_{\psi},S_{int}$ as in Eq.\eqref{Eq:model} and a bosonoic propagator $D(\mathbf{q},\Omega)= \delta(\mathbf{q-Q},\Omega)$. The fermions tend to an AFM order with  $\mathbf{Q}=\left(\pi,\pi \right)$. The self consistent (Eliashberg) equation for $\Sigma \left(\mathbf{k},i\omega \right)$ is given by
\begin{align}
\Sigma \left(\mathbf{k},i\omega \right) = \int_{k',\omega'} G \left(\mathbf{k'},i\omega' \right)  D \left(\mathbf{k-k'},i\omega-i\omega' \right) 
\end{align}
Using the above definition for $D$ and $G \left(\mathbf{k'},\omega' \right)= \left( i\omega'-\epsilon_{\mathbf{k'
}} -\Sigma (\mathbf{k}',\omega') \right ) ^{-1} $ , we can solve  for $\Sigma$ at the hot spot ($\mathbf{k}=0$). 
\begin{align}
\Sigma(0,i\omega) &=g^2 /(i\omega-\Sigma) \notag  \\
\Rightarrow \Sigma(0,i\omega)&= i\cdot\frac{\omega \pm \sqrt{\omega^2 +4g^2}}{2} .
 \end{align}
After analytical continuation, we find that
 \begin{align}
     A(0,\omega)=
     \begin{cases}
    \ \frac{\sqrt{4g^2 -\omega^2 }}{2\pi g^2} , & \text{if}   \ \left| \omega \right| \leq 2g  \\
        \ 0  ,  & \text{if} \ \left| \omega \right| \geq 2g . \\ 
  \end{cases}
 \end{align}
On the other hand, this simple case can be solved exactly. For a given configuration $\phi $, finding $G_{\phi} (\mathbf{k},i\omega)$ is straightforward. We find that at the hot spot 	
\begin{align}
G(0,i\omega)=\left( iI\omega+g \tensor{\sigma} \vec{\phi} \right) ^{-1}=\frac{ -iI\omega+g \tensor{\sigma} \vec{\phi} }{\omega^{2}+ g^{2} \left|\phi \right|^{2}} .
\end{align}
 We therefore  find the following spectral function (after analytical continuation, $i\omega \to \omega +i
\eta$)
\begin{align}
    A_{\phi}(0,\omega)=-\pi^{-1}\Im G_{\phi,\uparrow \uparrow} \left( 0, \omega \right)=\frac{1}{2} \left( \delta \left(\omega- g \left|\phi \right| \right)+\delta\left(\omega+g \left|\phi \right| \right) \right) .
\end{align}
Integrating out the field $\phi(\mathbf{q})$ at $\mathbf{q}=\mathbf{Q} $  yields
\begin{align}
   \frac{ \int  A(\mathbf{k_{HS}},\omega)  e^{-\phi^{2}}
\phi^{2} \sin{\theta} d\phi  d\theta d\varphi }{ \int   e^{-\phi^{2}}
\phi^{2} \sin{\theta} d\phi  d\theta d\varphi }= \frac{2\omega^2}{g^3\sqrt{\pi}}e^{-\frac{\omega^2}{g^2}}.
\end{align}
These two results are qualitatively different (see Fig \ref{fig: SC_p_Eliashberapg}). We therefore deduce that when having a bosonic propagator which is sharply peaked at $\mathbf{q}=\mathbf{Q},\Omega=0$, summing over the rainbow diagrams only (ignoring vertex corrections) is not sufficient (unlike the overdamped case, where the rainbow summation is known to give a good approximation for the self-energy).
\section{Simulations}
\label{appendix:simulation}
{In this section, we will review the main technical steps of the numerical calculations, and present an analysis of finite size and finite Trotter step effects}
\subsection{Numerical algorithm: single particle properties}
\label{appendix:simulation_process_GF}
Since our simulations were performed for $\gamma=0$, the bosons are not affected by the femionic determinant. Hence, we sample $\phi$ directly from a Gaussian distribution with zero expectation
and with the following standard deviation:
\begin{align}
\sigma= \left(2\left(  \xi^{-2} + 4\sin{^{2}\left(\frac{q_{x}}{2}\right)}+ 4\sin{^{2}\left(\frac{q_{y}}{2}\right)} +\frac{4\sin{^{2} \frac{\left(\Omega_n\Delta\tau\right)}{2}}}{\left(v_s\Delta\tau\right)^{2}} \right) \right)^{-\frac{1}{2}},
\end{align}
which originates from the lattice version of the continuous bosonic propagator, $D\left(\mathbf{q},i\Omega_n\right)$ (see Eq.(\ref{Eq:model})). Here, the lattice spacing is $a=1$ and  $\Delta\tau$  is the imaginary time step.

We can easily obtain $\phi(\mathbf{r},\tau)$ by a Fourier transform and use it for $H_{\rm{int}}$ (by the proper real space form of Eq.(\ref{Eq:model}) on a lattice).  
Our free Hamiltonian $H_{\psi}$ was written in real space, with chemical potential $\mu$ 
and tight binding nearest and next-nearest neighbours hopping amplitudes: $t_{A}, t_{B}$ respectively.
For a given system size $L^2$ and time steps $N_{\tau}=\beta/\Delta\tau$, the single-particle action matrix os of dimension $2L^2 N_{\tau}\ \times\ 2L^2 N_{\tau} $. The factor 2 is due to spin. 

The Green's function is then obtained by the matrix inversion
\begin{align}
(-\partial_{\tau}+H_{\psi}+H_{\rm{int}}) G_{\phi}(\tau,\mathbf{r},\tau',\mathbf{r'}) =    \delta(\mathbf{r-r'})\delta(\tau-\tau')
.\end{align}
Using translational symmetry (after the averaging $\left<G_{\phi}\right>$), it is enough to find $G \left( \mathbf{r+r'},\tau +\tau' ; \mathbf{r'},\tau' \right)$ for a given $\mathbf{r}'$. We can therefore solve a system of equations (for a specific column $\mathbf{r'},\tau'$ )  rather than inverting the entire matrix.

Having found the Green's function, we can find $\Im\Sigma(\mathbf{k}\approx\mathbf{{k}_{HS}},i\omega)$, shown in Fig.\ref{fig: simulation_Sigma}.
\\
\\
The quasi-particle weight at $\omega=0$ is then 
obtained from $Z=1/\left(1-\left. \frac{\partial \Im\Sigma(\mathbf{k},i\omega)}{\partial \omega}\right\vert_{\omega=0}\right)$ . We estimate the derivative  $\left. \frac{\partial \Im\Sigma(\mathbf{k},i\omega)}{\partial \omega}\right\vert_{\omega=0}$ by using the anti-symmetric property of $\Im\Sigma(\mathbf{k},i\omega)$ as follow: We take the smallest Matsubara frequencies $(\pm \pi T , \pm  3\pi T , \pm 5\pi T )$
and perform a polynomial fit $ax^3 +bx$. The coeffcient $b$ is our estimate for   $\left. \frac{\partial \Im\Sigma(\mathbf{k},i\omega)}{\partial \omega}\right\vert_{\omega=0}$ .
\subsection{Superconducting susceptibility}
\label{appendix:simulation_process_sc}

In Sec.~\ref{SC:results}, we present our results for the superconducting susceptibility. Technical details of the calculation, including finite size and finite time step effects, are discussed in Sec.~\ref{SC:details}.

\subsubsection{results}
\label{SC:results}
 
 Fig.~\ref{fig: simulation_SC} shows the results for the d-wave and s-wave susceptibilities, defined as:
\begin{figure}
        	\centering\includegraphics[ width=0.7\columnwidth]{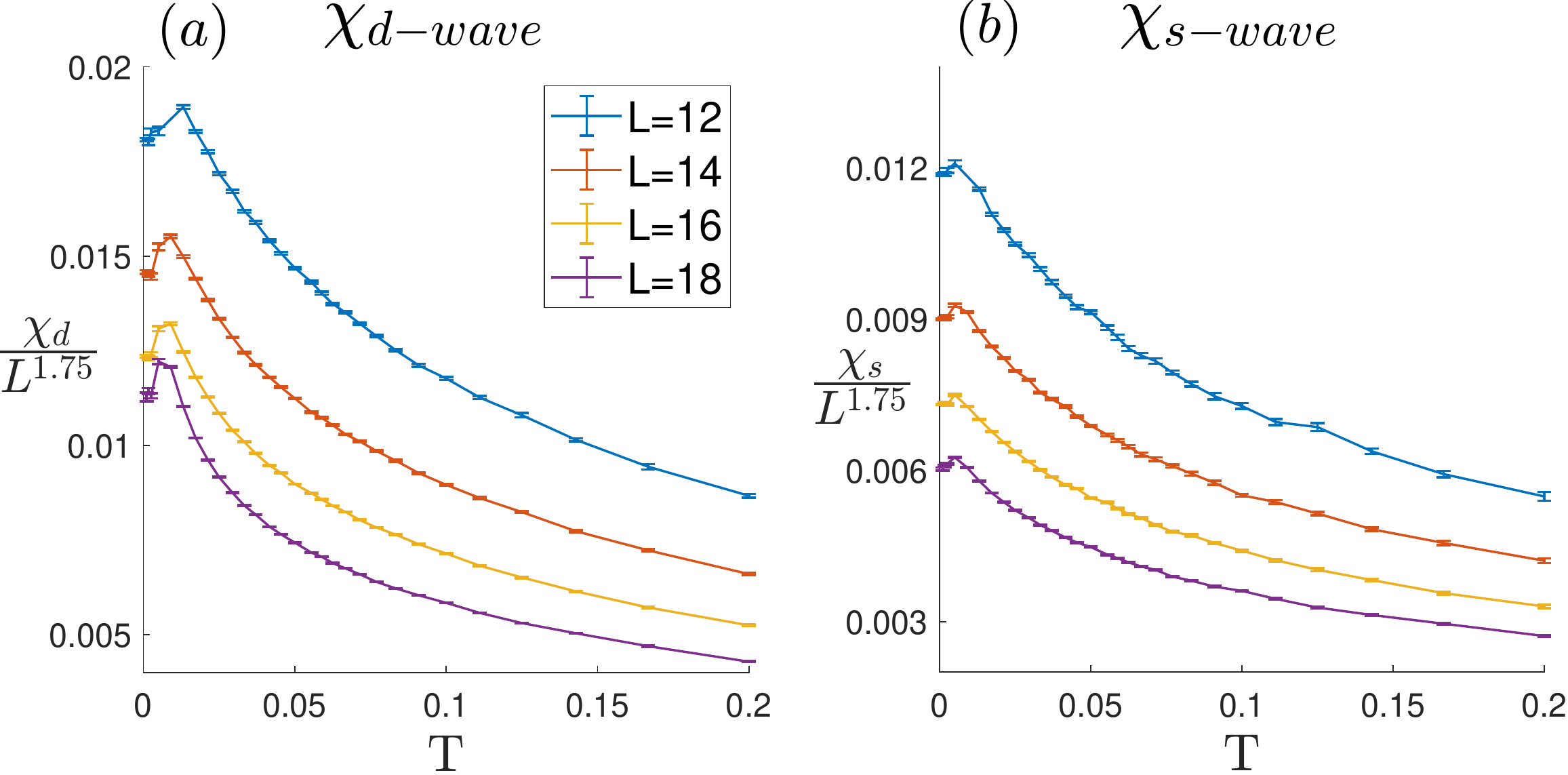}
	\caption{The results of the paring susceptibility, divided by $L^{1.75}$,  for various system sizes and:  $0.001<T<0.2$  (\textbf{a}) for d-wave superconductivity  (\textbf{b}) for s-wave superconductivity. }
\label{fig: simulation_SC}
\end{figure}
\begin{align}
\chi_{s,d} &= \int^{\beta}_{0} d\tau \sum_{i} \left<  \Delta^{\dagger}_{s,d}  \left( \mathbf{r}_i ,\tau \right)  \Delta_{s,d} \left( 0,0 \right) \right>,
\label{Eq: chi_SC_def}
 \end{align}  
  where $ 
    \Delta_{d}  \left( \mathbf{r}_i \right)=\sum_{j}\eta_{i,j} \left( \psi_{i,\uparrow} \psi_{j,\downarrow} -  \psi_{i,\downarrow} \psi_{j,\uparrow} \right) $, and
$   \Delta_{s}  \left( \mathbf{r}_i    \right)=\psi_{i,\uparrow}\psi_{i,\downarrow}$. Here, $\eta_{i,j}=\frac{1}{4}$ if $\mathbf{r}_i - \mathbf{r}_j =\pm\hat{\mathbf{x}}$, $\eta_{i,j}=-\frac{1}{4}$ if $\mathbf{r}_i - \mathbf{r}_j =\pm\hat{\mathbf{y}}$ and $\eta_{i,j}=0$ otherwise. The susceptibilities are scaled by $L^{1.75}$, where $L$ is the linear system size. We find no intersection between the different system sizes; moreover, the superconducting susceptibility does not exhibit a strong enhancement at low temperature, which may indicate the present of significant superconducting fluctuations. We therefore conclude that the observation of $Z \sim 1/\xi$ may be attributed to AFM fluctuations, as the perturbative calculation suggests.

\subsubsection{Technical details of the calculation of the SC susceptibilities}
\label{SC:details}
It is well known that inserting an artificial orbital magnetic field that vanishes in the thermodynamic limit can dramatically improve the convergence with system size\cite{Assad2002}. 
However, reading single particle properties when introducing flux becomes problematic. When a magnetic flux is included, the operator $c^{\dagger}_{i,\sigma} c_{j,\sigma}$ is not gauge invariant and is therefore not physical. On the other hand, operators such as the superconducting susceptibility $\chi_{s},\chi_{d}$, can be defined in a gauge invariant way even in the presence 
of magnetic flux. It is convenient to introduce the fictitious magnetic field such that it couples with an opposite sign to spin up and spin down electrons. The flux is implemented by a Peiels substitution, adding a phase to the hopping amplitudes - such that we accumulate a phase of $2\pi/(L^2)$ along the orbit of one unit cell. For the purpose of calculating the superconducting susceptibility, we choose the magnetic flux to couple with an opposite sign to spin up and spin down electrons, such that spin-singlet Cooper pairs are not subject to the flux. The process of choosing the proper phases $A_{i,j}$ for $t_{ij,\sigma}e^{A_{i,j}^{\sigma}}$ on each bond is explained in details in \cite{Berg2015}. We restrict ourselves to a subclass of gauges such that $A_{\uparrow}=-A_{\downarrow}$. Hence the hopping amplitudes satisfy $t_{ij,\uparrow}=t^{*}_{ij,\downarrow} $.

We define the $d$-wave susceptibility $\chi_{d}$ in the presence of flux as:
\begin{align}
     \Delta_{d}  \left( \mathbf{r}_i \right)=\sum_{j}\eta_{i,j} \left( \psi_{i,\uparrow} \psi_{j,\downarrow} e^{i A_{i,j}^{\downarrow}} -  \psi_{i,\downarrow}  \psi_{j,\uparrow} 
     e^{i A_{i,j}^{\uparrow}}
     \right) .
     \label{Eq:flux_chi_d_def}
\end{align}
 In the thermodynamic limit, $A_{i,j}^\sigma \rightarrow 0$ (since there is only one flux quantum in the entire system), and Eq.(\ref{Eq:flux_chi_d_def}) coincides with Eq.\eqref{Eq: chi_SC_def}. Following the same logic for $\chi_{s}$ leads to the original definition with no dependence in $A_{i,j}$ (as in Eq.\eqref{Eq: chi_SC_def}) due to cancellation of phases as the pairing is onsite. Hence for $\chi_{s}$ we can stick to the definition in Eq.(\ref{Eq: chi_SC_def}) without flux.

In addition, we have to consider the effect of flux on the field $\vec{\phi}$. Note that, since the flux couples to the z-components of the spin (i.e., it couples oppositely to electrons with opposite spin), the bosonic action needs to be modified such that it is gauge invariant. The appropriate continuum form of the bosonic action in the presence of a flux is
\begin{align}
S_{\phi}&=  \int d\tau d\mathbf{r} \left[ 
    \frac{1}{v_{s}^2}\left( \partial_{\tau} \vec{\phi}^{2} \right) 
  + \left|  
  \left(  \grad -2i A^{\uparrow}(\mathbf{r})\sigma^{y} \right) \begin{pmatrix}
\phi_x\\
\phi_y 
\end{pmatrix}  \right|^{2}
   +   \left(  \grad \phi_z \right)^2+ \xi^{-2} 
    \right] .
    \label{Eq:model_phi_with_flux}   
\end{align}

Then, the entire action is invariant under spin-dependent gauge transformations of the form: 
  \begin{align}
    \begin{cases}
   \begin{pmatrix}
\psi_{\mathbf{r},\uparrow} \\
\psi_{\mathbf{r},\downarrow} 
\end{pmatrix} \to e^{i\alpha(\mathbf{r})\sigma_z} \begin{pmatrix}
\psi_{\mathbf{r},\uparrow} \\
\psi_{\mathbf{r},\downarrow} 
\end{pmatrix} \\
   A^{\uparrow}(\mathbf{r}) \to A^{\uparrow}(\mathbf{r}) +\grad{\alpha(\mathbf{r})}\\
   A^{\downarrow}(\mathbf{r}) \to A^{\downarrow}(\mathbf{r}) -\grad{\alpha(\mathbf{r})}\\
   \begin{pmatrix}
\phi_x(\mathbf{r})\\
\phi_y (\mathbf{r})
\end{pmatrix} \to e^{2i\alpha(\mathbf{r})\sigma_y} \begin{pmatrix}
\phi_x(\mathbf{r})\\
\phi_y (\mathbf{r})
\end{pmatrix} \\
   \phi_z(\mathbf{r}) \to  \phi_z(\mathbf{r})
    \end{cases}
    \label{Eq: gauge_transform},
    \end{align} 
Note that in Eq.\eqref{Eq:model_phi_with_flux} , the vector potential does not couple to $\phi_z$, since this component is invariant under spin rotations around the $z$ axis.
The lattice version of the action can be obtained in a similar manner when considering the discrete version of the bosonic action. 

\subsection{Finite size and finite imaginary time step effects}
\label{appendix:size_effects}
\begin{figure}
        \includegraphics[ width=0.7\columnwidth ]{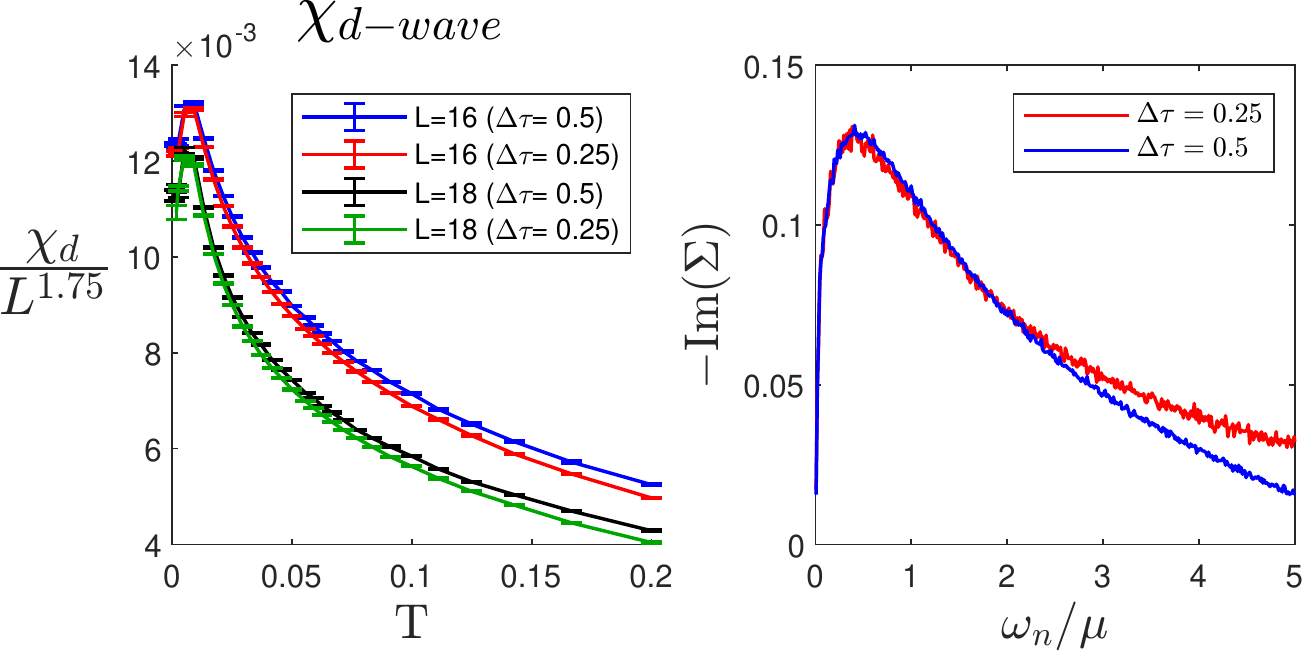}
	\caption{Size effects in imaginary time for the d-wave susceptibility (\textbf{a}) and for the self energy (\textbf{b}) , for the maximal $\xi$ value ($\xi\Lambda\approx 16$) at $T=0.002$. It is clear that at the limit $T/\mu,\omega_n/\mu \ll 1$ , it is sufficient to have  $\Delta\tau=0.5$ in order to avoid substantial size effects.   }
\label{fig: dtau_size_effects}
\end{figure}

\begin{figure}
 	\hspace{-0.18cm}
      	\centering\includegraphics[ width=0.7\columnwidth ]{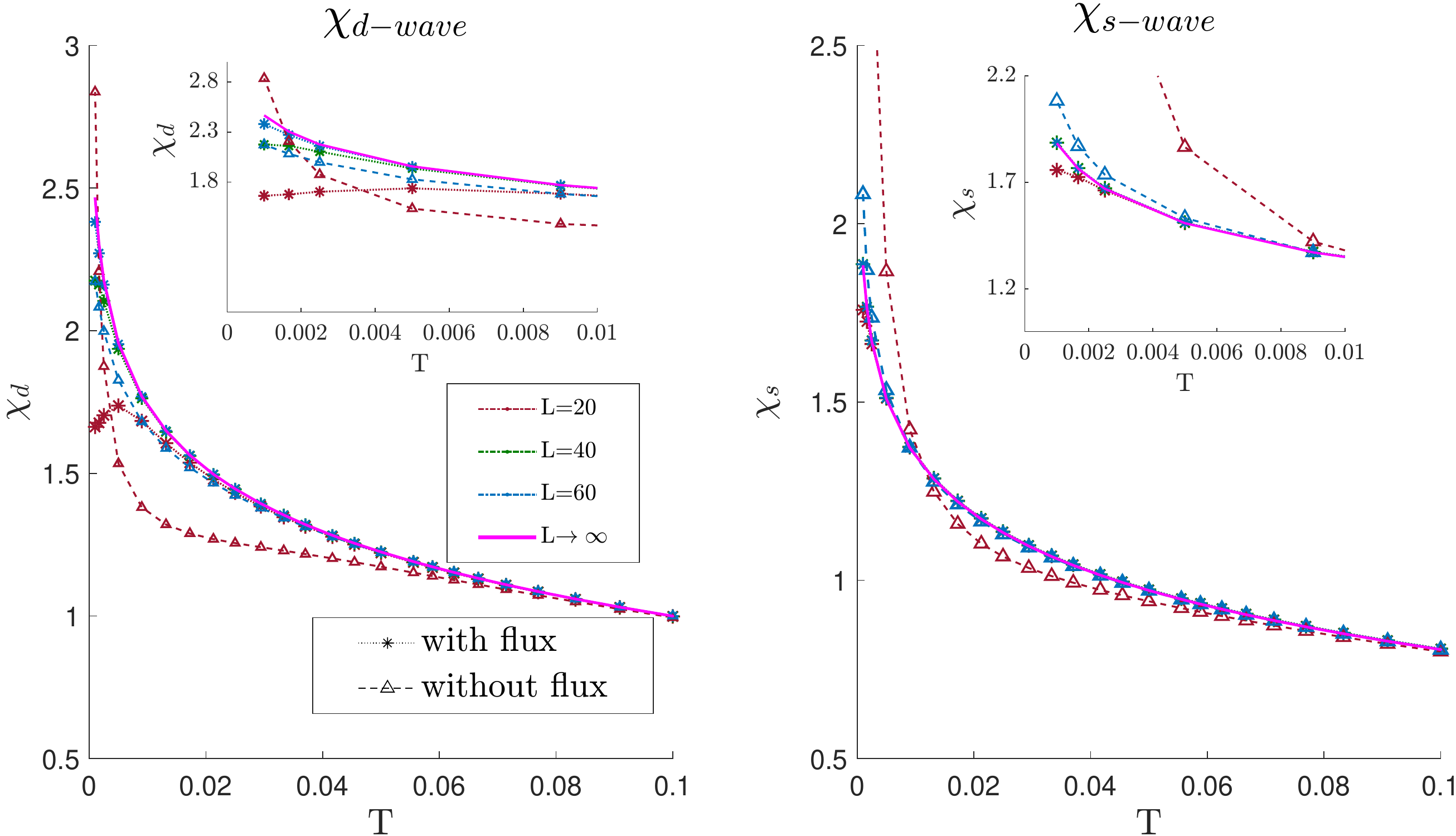}
	\caption{(\textbf{a}) $\chi_d$  and  (\textbf{b}) $\chi_s$  with no interactions (as in the main text  $t_A=-0.85\mu,t_B=-0.45\mu$). The results with flux (marked by $*$) converge much quicker to the thermodynamic limit (solid line) in $\chi_s$ than in $\chi_d$, but in both cases they are significantly better than the  results without flux (marked by $\Delta$).  }
\label{fig:SC_comparison_no_int}
\end{figure}

In our simulations, we have used imaginary time steps of $\mu\Delta\tau=0.5$. Although this is a relatively large time step, it is sufficiently small when studying the behaviour at small frequencies. The results for $\chi_d$ and $\mathrm{Im}\Sigma(\mathbf{k}\approx \mathbf{{k}_{HS}},i\omega_n)$ are shown in Fig.~\ref{fig: dtau_size_effects} for two values of $\Delta\tau$ ($\chi_d$ was computed with flux, according to Eq.\eqref{Eq:flux_chi_d_def}).
We find that for the lowest temperatures and frequencies, the results are essentially independent of $\Delta\tau$. 
At low $T$, the SC susceptibility exhibits a non-monotonic behavior (Fig.~\ref{fig: simulation_SC}). Both $\chi_d(T)$ and $\chi_s(T)$ show a maximum at finite $T$; in $\chi_d(T)$, it is apparent that the maximum shifts towards lower temperature as the system size increases .We believe that this non-monotonic behavior is a finite size effects in the presence of a flux. To demonstrate this, we have calculated the finite size non-interacting superconducting susceptibility, with and without flux (see Fig.~\ref{fig:SC_comparison_no_int}). The non-monotonic behavior is visible in $\chi_d(T)$ without interactions; nevertheless, at temperatures above the maximum in $\chi_d(T)$, the results for systems with flux converge much more rapidly to the thermodynamic limit than those without flux.  

\begin{figure}
        \includegraphics[ width=0.55\columnwidth ]{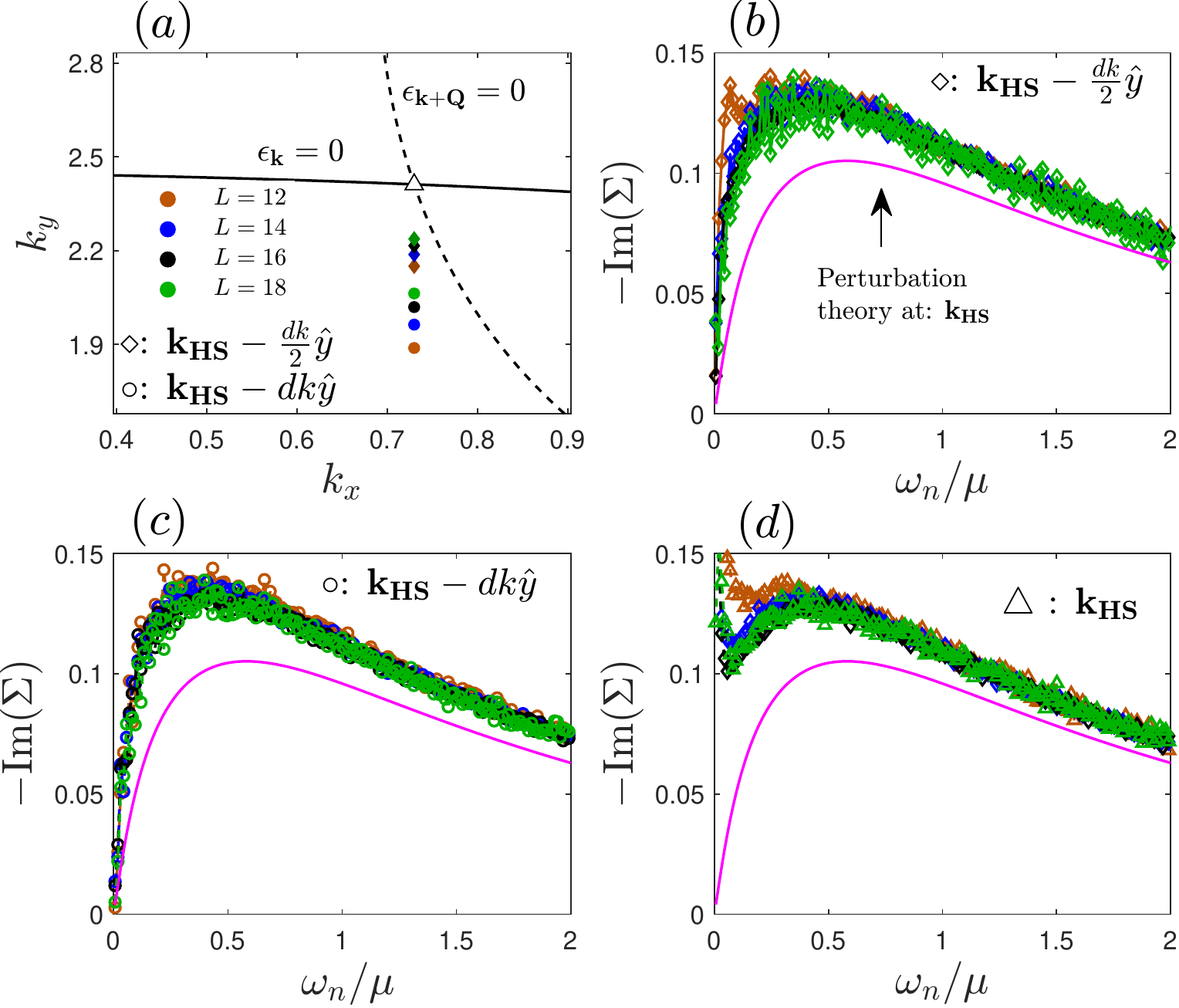}
	\caption{Size effects for different system sizes, at  $\Lambda\xi=16,t_A=-0.85\mu,t_B=-0.45\mu,T=0.002$. Close to $\mathbf{k_{HS}}$  (\textbf{b-c}, according to the markers and colors in panel \textbf{a} of the BZ), and exactly at the hot spot  (\textbf{d}).  It is clear that close to the hot spot we produce qualitatively similar results to those in perturbation theory. However a clear upturn was observed exactly at the hot spot.   }
\label{fig: spatial_size_effects}
\end{figure}

In addition, in Fig.~\ref{fig: spatial_size_effects}, we also show $-\rm{Im}\Sigma$ with $\mu\Delta\tau=0.5$ for different system sizes: $L=12,14,16,18$.
We can design our momentum grid by using twisted boundary conditions (TBC). In particular, we use TBC such that the grid includes one hot spot pair for every system size. From the lattice dispersion $
\epsilon_{\mathbf{k}}=2t_{A} \left( \cos k_x+\cos k_y \right)   
+4t_{B}\cos{k_x}\cos{k_y}-\mu$, The hot spots can be found analytically. We easily find that one of the points is $\mathbf{k}_{\mathbf{HS},x} =\frac{1}{2}\left(\pi +\cos{^{-1} \left(1+\frac{\mu}{2t_{B}}\right)} \right)  $ and $\mathbf{k}_{\mathbf{HS},y}=\pi- \mathbf{k}_{\mathbf{HS},x}$.

When the momentum grid includes a point at the hot spot (Fig.~\ref{fig: spatial_size_effects}(d)), $-\rm{\Sigma}(\mathbf{{k}_{HS}},i\omega)$ exhibits a low frequency upturn at a system size-dependent $\omega$. 
It is not clear if this behavior survives in the thermodynamic limit. 
To test this, we have examined $-\rm{Im}\Sigma(\mathbf{k}\approx \mathbf{{k}_{HS}},i\omega_n)$ at a point near the hot spot, at $\mathbf{k}=\mathbf{k_{HS}}-dk\, \hat{y}$ with $dk=2\pi/L$. In addition, we have used an alternative boundary conditions such that the grid point closest to the hot spot is at $\mathbf{k}=\mathbf{k_{HS}}-dk/2 \hat{y}$. The results for $-\rm{Im}\Sigma(\mathbf{k}\approx \mathbf{{k}_{HS}},i\omega)$ near the hot spot for these two choices of the boundary conditions are presented in Fig.~\ref{fig: spatial_size_effects}(b) and (c). As can be seen from the figure, near the hot spot, $-\rm{Im}\Sigma(\mathbf{k}\approx \mathbf{{k}_{HS}},i\omega)$ is weakly size dependent. The low-frequency upturn is absent, and the behavior of $-\rm{Im}\Sigma(\mathbf{k}\approx \mathbf{{k}_{HS}},i\omega)$ is qualitatively similar to the behavior expected from perturbation theory (solid line). We therefore conclude that the low-frequency upturn in $-\rm{Im}\Sigma(\mathbf{k}=\mathbf{{k}_{HS}},i\omega)$ is likely to be a finite size effect, and the behavior seen in Fig.~\ref{fig: spatial_size_effects}(b,c) is more representative of the $L\rightarrow \infty$ limit. 

\clearpage
\end{document}